\documentclass[lettersize,journal]{IEEEtran}
\usepackage{amsmath,amsfonts}
\usepackage{array}
\usepackage[caption=false,font=footnotesize,labelfont=rm,textfont=rm]{subfig}
\usepackage{textcomp}
\usepackage{stfloats}
\usepackage{url}
\usepackage{verbatim}
\usepackage{graphicx}
\usepackage{cite}
\usepackage{epstopdf}  
\usepackage{float}
\usepackage{amssymb}
\usepackage{color}
\usepackage{enumerate}
\usepackage{wrapfig}

\usepackage[ruled]{algorithm2e} 

\SetAlFnt{\small}
\SetAlCapFnt{\small}
\SetAlCapNameFnt{\small}
\SetAlCapHSkip{0pt}

\def\0{\mathbf{0}}
\def\1{\mathbf{1}}

\def\ba{\mathbf{a}}
\def\bb{\mathbf{b}} 

 \def\bD{\mathbf{D}}
\def\be{\mathbf{e}} 
\def\bff{\mathbf{f}}
\def\bg{\mathbf{g}} 
 
 \def\bI{\mathbf{I}} 
\def\bJ{\mathbf{J}}
\def\bk{\mathbf{k}} \def\bK{\mathbf{K}}

\def\bM{\mathbf{M}} 
 \def\bN{\mathbf{N}}

\def\bq{\mathbf{q}} \def\bQ{\mathbf{Q}}
\def\br{\mathbf{r}}

\def\bu{\mathbf{u}} 
\def\bv{\mathbf{v}} 
 
\def\bx{\mathbf{x}} \def\bX{\mathbf{X}}

\newcommand{\bpsi}{\mbox{\boldmath$\psi$}}
\newcommand{\bPsi}{\mbox{\boldmath$\Psi$}}
\newcommand{\bPhi}{\mbox{\boldmath$\Phi$}}

\newcommand{\btau}{\mbox{\boldmath$\tau$}}

\def\calV{\mathcal{V}}
\def\calX{\mathcal{X}}

\def\rd{\mathrm{d}}

\usepackage{leftidx}

\def\eb{\begin{equation}}
\def\ee{\end{equation}}

\usepackage{makecell}

\begin{document}
\title{Explicit Topology Optimization of Conforming Voronoi Foams} 

\author{
    Ming Li, Jingqiao Hu, Wei Chen, Weipeng Kong, and Jin Huang
\thanks{The authors are with State Key Laboratory of CAD\&CG, Zhejiang University, Hangzhou, 310027, China. E-mail: \{liming, hj\}@cad.zju.edu.cn, \{hujingqiao, chen\_w, wpkong\}@zju.edu.cn.}
}



\maketitle

\begin{abstract}
  Topology optimization is able to maximally leverage the high DOFs and mechanical potentiality of porous foams but faces three fundamental challenges: conforming to  free-form outer shapes, maintaining geometric connectivity between adjacent cells, and achieving high simulation accuracy. To resolve the issues, borrowing the concept from Voronoi tessellation, we propose to use the site (or seed) positions and radii of the beams as the DOFs for open-cell foam design.  Such DOFs cover extensive design space and have clear geometrical meaning, which makes it easy to provide explicit controls (e.g. granularity). During the gradient-based optimization, the foam topology can change freely, and some seeds may even be pushed out of the shape, which greatly alleviates the challenges of prescribing a fixed underlying grid. The mechanical property of our foam is computed from its highly heterogeneous density field counterpart discretized on a background mesh, with a much improved accuracy via a new material-aware numerical coarsening method. We also explore the differentiability of the open-cell Voronoi foams w.r.t. its seed locations, and propose a local finite difference method to estimate the derivatives efficiently. We do not only show the improved foam performance of our Voronoi foam in comparison with classical topology optimization approaches, but also demonstrate its advantages in various settings, especially when the target volume fraction is extremely low. 
\end{abstract}

%
%
  
%
%


\begin{figure*}[!htb]
  \centering
  \includegraphics[width=0.8\textwidth]{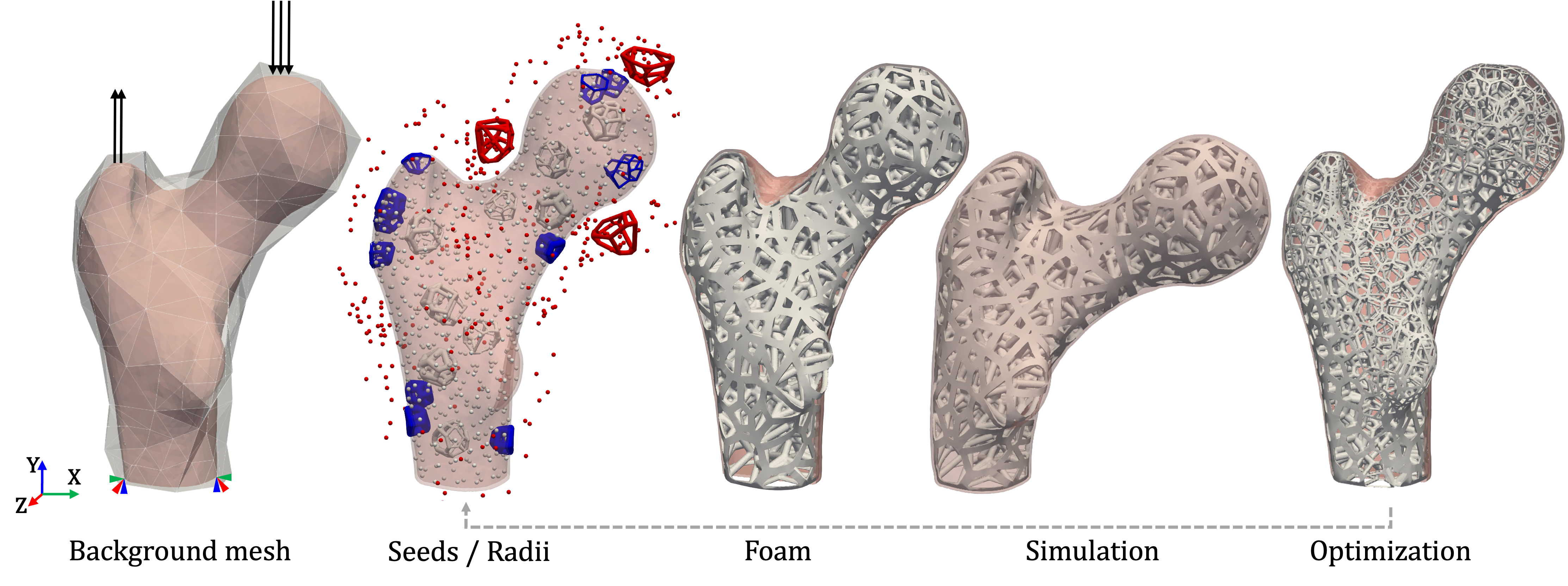} 
  \caption{We optimize the topology of open-cell Voronoi foams taking the Voronoi seeds and edge radii as design variables. It for the first time achieves a gradient-based topology optimization of conforming foams with an ensured valid geometry connections.  
We carefully approximate the derivatives of Voronoi diagram and coarsen the equilibrium equation by material-aware polyhedral shape functions, and make the gradient-based optimization highly efficient.}
  \label{fig:teaser}
\end{figure*}



%
%


\section{Introduction}\label{sec-introduction} 
\IEEEPARstart{P}{orous} foams have attractive and distinguishing properties of lightweight, high stiffness-ratio, energy absorption and flexibly tailored rigidity and so on~\cite{app10186374, fernandes2021mechanically, Zhang2017Nature,Lesson2022}.
Topology optimization is very effective in maximally leveraging the high DOFs and mechanical potentiality of porous foams~\cite{Wu2021Review,Liu2021MLattice}.  
The prominent approaches generally assume that specific porous cells are distributed within a prescribed axis-aligned regular grid. This simplifies modeling, simulation and optimization, but also restricts the solution space, and the achievable structural performance.  Especially, the following three requirements bring big challenges to existing techniques~\cite{Wu2021Review,Liu2021MLattice}: free-form conforming, geometric connectivity and simulation accuracy. 

This important topic, which has not yet been fully noticed or carefully addressed, are to be explored in this study. We particularly focus on open-cell foams as their open interior allows for ease of material clearance after fabrication. We set up an explicit design space for open-cell foam with clear topology and geometry control parameters, borrowing the concept from Voronoi tessellation~\cite{du1999centroidal}, and develop an \emph{explicit topology optimization} approach to optimize its performance; see Fig.~\ref{fig:teaser} for an illustration. 
Actually, natural open-cell foams are often idealized as edges of Voronoi cells ~\cite{1997Cellular, martinez2016procedural}. Utilizing Voronoi foams have attracted a lot attentions in computer graphics and mechanical engineering~\cite{2020Investigation, martinez2016procedural, martinez2017orthotropic, martinez2018polyhedral,podesta2018material}.
However, topology optimization of open-cell foams are rarely studied. Most methods just directly specify the number (or density) of the seeds, even their locations, to design the structure, and lack the ability of optimizing it w.r.t. mechanical goal in a variational way. In addition, simulation of high accuracy was rarely studied. 


The proposed approach is applicable to a wide range of physics-based porous foam optimization. We specifically focus on a widely studied stiffness maximization for proof of concept, and also for a representative performance comparisons. The study has the following main contributions. 
\begin{enumerate}
\item Being able to simultaneously optimize a foam's topology and geometry under explicit control parameters with extensive design space. The approach is also able to tune the foam cell number automatically, which was seldom observed in previous studies of biscale-topology optimization or cell-tiling-based optimization.
\item A synchronized explicit and implicit foam representations for topology optimization of conforming Voronoi foams. It forms a seamless pipeline that conducts the modeling, simulation, gradient computation on a uniform implicit representation, avoiding the unstable and time-consuming model conversions between modeling and simulation. 
\item A numerical coarsening approach to simulate the deformation of the complicated porous foams. It solves the equilibrium equation about its high resolution heterogeneous density field without assumption of scale separation, and reduces the computational costs by an order of magnitude compared with benchmark FEM results.  
\item An approach to deeply explore the differentiability of 3D open-cell Voronoi foams w.r.t. to its seed locations, and the associated gradient-based topology optimization framework. The gradient is computed via a local finite difference approach, without efforts of expensive Voronoi construction, and significantly improves the efficiency.
\end{enumerate}

The remainder of the study is arranged as follows. Related work is discussed in Section~\ref{sec-related}. The idea of Voronoi foam design is explained in Section~\ref{sec-problem}. Techniques on design space definition, design goal formulations, optimization and simulation method are respectively explained in Sections~\ref{sec-space}, \ref{sec:goal}, ~\ref{sec:opt} and \ref{sec-coarsening}. Extensive numerical examples are demonstrated in Section~\ref{sec-examples}, followed by the conclusion in Section~\ref{sec-conclusion}. 

\section{Related work}\label{sec-related}
We discuss the related work on simulation, optimization of porous foams and their conforming design. 

\subsection{Foam simulations} \label{sec-related-sim}
Property of a porous foam can be directly predicted via using classical FE method via tessellating it into a discrete volume mesh. However, its too complex geometric structure poses severe challenges on its reliable FE mesh generation and efficient solution computation. 

Typically, a foam is simulated via numerical homogenization~\cite{bensoussan1978asymptotic,sigmund1994materials}. It is achieved via two levels of FE computations - coarse-level and fine-level wherein the simulation results on each foam cell are used in parallel to predict the overall performance in the coarse-level, and vice versa. The approach replaces each foam cell with an effective elasticity tensor via the asymptotic~\cite{pinho2009asymptotic,chung2001asymptotic,andreassen2014determine} or energy-based approximations~\cite{sigmund1994materials}, at an assumption of scale separation and periodic cell distribution. In contrast, the Voronoi foams studied here have cutout or deformed cells or full-solid covering shells, which seriously breaks the assumption. Applying directly these approaches may much reduce the simulation accuracy. 
The reduced order model (ROM) was recently proposed to simulate the porous foams~\cite{White2023ARO} based on previous studies~\cite{Eftang2013PortRI,McBane2020ComponentwiseRO}. It shares the same spirit with the approach in representing the shape function as a matrix transformation~\cite{li2022analysis}. 
%
Note that it is not very reasonable to simulate our foam as an assembly of \emph{beam elements}~\cite{bower2009applied} as it involves smooth blends between the beams.  

Being embedded within a coarse background mesh, a porous foam can be taken as a heterogeneous structure and simulated via numerical coarsening with no assumption of scale separation~\cite{kharevych2009numerical,chen2015data}. We here construct material-aware shape (or ``basis'') functions to reflect finely the material distribution within each coarse element, which has shown great potential in improving the simulation efficiency and accuracy of heterogenous structures~\cite{nesme2009preserving,torres2016high,chen2018numerical}. Piecewise-trilinear shape functions were initially introduced ~\cite{nesme2009preserving}. Later on matrix-valued form was devised to capture the non-linear stress-strain behavior with an improved simulation accuracy~\cite{chen2018numerical}, achieved via solving a relatively expensive optimization problem. Very recently, the shape functions in an explicit form of matrix product were introduced~\cite{li2022analysis}, which overcomes the challenging issue of inter-element stiffness and ensures the fine-mesh solution continuity. We further extend the approach in this study to simulate Voronoi foams on general background polyhedral meshes. The approaches share a similar spirit with finite cell method (FCM)~\cite{schillinger2015finite,longva2020higher} using higher-order FE shape functions on an embedded mesh.

\subsection{Foam optimizations} 
Optimization of porous foams has been widely studied via topology optimization or parametric optimization~\cite{Wu2021Review,Liu2021MLattice}. The topology optimization is conducted separately in a single scale or concurrently in a biscale optimization~
\cite{schumacher2015microstructures,zhu2017two,liu2018narrow,Wu:2017,wu2019design}. When in a single scale, it requires imposing local volume constraints to generate bone-like foams~\cite{Wu:2017}, or constraints of solid and void sizes~\cite{Fernandez2020Size,zhu2020Areview_40}. These constraints were tediously designed manually, and raised additional difficulty in its optimal solution computation. When in biscale, maintaining the geometric connection between adjacent cells comes out as a fundamental challenge. Huge research have been devoted to resolve them~\cite{schumacher2015microstructures,garner2019compatibility,hu2020cellular,zong2019vcut}, but they have no effective shape control ability due to the intrinsic low-level voxel-based shape representations. Geometrically invalid structures are usually found in the optimized structures that may contain broken, slender or small-void regions; see the examples in Section~\ref{sec-example-evf}. The recent approach of MMC (Moving Morphing Components)~\cite{RN233,Guo2017Self,du2022efficient} conducts the topology optimization using simple geometric primitives, which shares the same spirit of the study. 

Utilizing fixed type of parametric cells for form design optimization has also been studied. The foam cell parameter distribution was usually optimized for improved performance or to follow certain material properties or stress directions~\cite{panetta2015elastic,panetta2017worst,wu2019topology,Wang2013TOGCost,li2020anisotropic}. Various types of cells were explored, for planar rod networks~\cite{Schumacher2018MechanicalCO}, for strongly controlled anisotropy~\cite{Tricard2020FreelyOM}, or based on TPMS (Triply Periodic Minimal Surfaces) \cite{2020Efficient,Yan2020Strong3P}. \cite{tozoni2020low} optimized rhombic family for irregular foams conforming to an arbitrary outer shape. These approaches have ease in generating foams of valid geometry, at a cost of limited design choices. Great research efforts have been devoted to extend the cell types via de-homogenization~\cite{sigmund1994materials,panetta2015elastic, schumacher2015microstructures,chen2018computational}. 

\subsection{Conforming foam design}
Most of the above approaches work on a regular grid within a biscale framework linking the design domain and the foam cells. Their extensions to free-form shapes are generally achieved in three different ways: (1) assuming a sufficiently fine cell sizes and ignoring the cell-shape gap; (2) via boolean operations which may destroy the integrity of the boundary cells~\cite{bb_Cai2015Stress,Yoo2011}; (3) deforming foam cells to fit in a conforming hexahedral mesh~\cite{wang2005hybrid,tozoni2020low, Hong2021,Gupta2019}, where the cells may be extremely deformed. Recently, \cite{wu2019design} proposed an excellent approach of conforming foam design in two consecutive steps of material optimization and conforming foam generation. TPMS, as a special type of foams, has a merit in naturally maintaining the geometric validity after boolean operations with outer shapes~\cite{Yoo2011,2020Efficient,Yan2020Strong3P,2020Design}, but is limited in its restricted topological options. \cite{liu2021memory,ding2021stl} also presented a memory-efficient implicit representation of the foams.
 
Designing a conforming open-cell foam from Voronoi tessellations is very promising for its highly flexible topology and natural edge connectivity~\cite{du1999centroidal}. They were mostly achieved by aligning with a pre-optimized density, stress or material fields~\cite{martinez2016procedural, martinez2017orthotropic,wu2019design, Do2021Homogenization,Liu2021ADM, Liu2023Multiscale}. Most of them focused on 2D or 2.5D case~\cite{Do2021Homogenization,Liu2021ADM, Liu2023Multiscale}. An efficient 3D procedural modeling process~\cite{martinez2016procedural, martinez2017orthotropic} were introduced for open-cell foam construction with smoothly graded properties. 

A variational optimization will improve the convergence and the resulted foam performance. \cite{lu2014build} achieved this with maximal cell hollowing and minimum stress by an adaptive Monte Carlo optimization approach. Its huge computational costs and closed cells limit the industrial applications. Very recently, \cite{feng2022cellular} proposed an attractive concept of differentiable Voronoi diagrams via a continuous distance field approximation. The approach demonstrated its high efficiency, nice ability in anisotropy and locality control. It however only produced close-cell foams in 3D. \cite{Rakotosaona2021DST} studied the differential property of Voronoi tesselation restricted on a surface but it did not consider the physical property of the shape. 
 
Note again almost all the above approaches simulate the foam property via classical numerical homogenization which has a large accuracy loss; see also Section~\ref{sec-related-sim}. The issue is addressed here via a novel numerical coarsening approach.  

\section{Overview of the Algorithm}
\label{sec-problem}
To use Voronoi tessellations for open-cell foam design (see Fig.~\ref{fig:teaser}), one must carefully address the following three challenges:
\begin{itemize}
\item How to simulate the mechanical behavior of a Voronoi foam composed of many slender beams accurately. 
\item How to compute the derivatives about the edges (i.e. beams) in the Voronoi diagram w.r.t the design variables efficiently.
\item How to efficiently adapt the key parameters like cell number and foam topology under various constraints reliably. 
\end{itemize} 

Three technical points are proposed for the challenges: a synchronized explicit and implicit form for modelling a Voronoi foam, a numerical coarsening approach for simulation, and an efficient local finite difference approach to compute the gradients to guide the optimization. We coin the framework as the concept of \emph{explicit topology optimization}. Some technical considerations are explained below.

For explicit representation, we use the edges of Voronoi diagram with certain radii to form a Voronoi foam. The Voronoi seeds and the radii are called \emph{geometry-based design variables}. Approximating the structure with an implicit function produces a smooth Voronoi foam under explicit geometry-based variables. The representation allows for intricate structure control even at extremely low volume fraction, compared with the voxel-based or cell-based representation. The implicit representation avoids unreliable and time-consuming geometry conversion from modeling geometry to simulation mesh and allows for continuous FEM integration of high accuracy. 

The simulation of the Voronoi foams during iterative optimization has to be conducted on a fixed background mesh for reliable convergence, besides the general key properties of simulation efficiency and accuracy. A novel simulation strategy combining embedded simulation and numerical coarsening is proposed. Given a free-form outer shape, we construct a simple and coarse background mesh to embed it, avoiding error-prone boundary conforming mesh generation and reducing the computational costs. A set of material-aware shape functions for each coarse element can then be derived that tremendously reduces the simulation costs. 

We then formulate a variational optimization problem w.r.t the geometry-based design variables to optimize the foam's stiffness along with volume constraints and shape regularization requirements. 
For efficient gradient computation, we utilize the locality property of the Voronoi tessellations, and further propose a three-point based distance approximation approach. The overall process much alleviates the computational efforts by several orders of magnitude. 

\section{Design space of Voronoi foams}\label{sec-space}
\begin{figure}[!htb]
  \centering
  \subfloat[Multiple beams]{\includegraphics[width=0.14\textwidth]{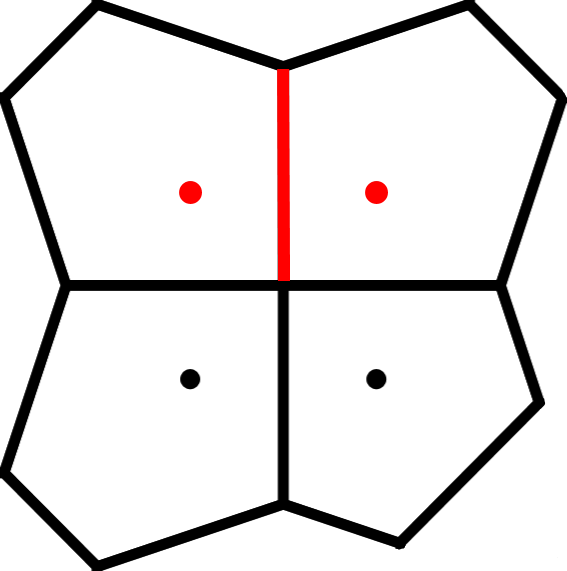}}\qquad 
  \subfloat[Single beam]{\includegraphics[width=0.18\textwidth]{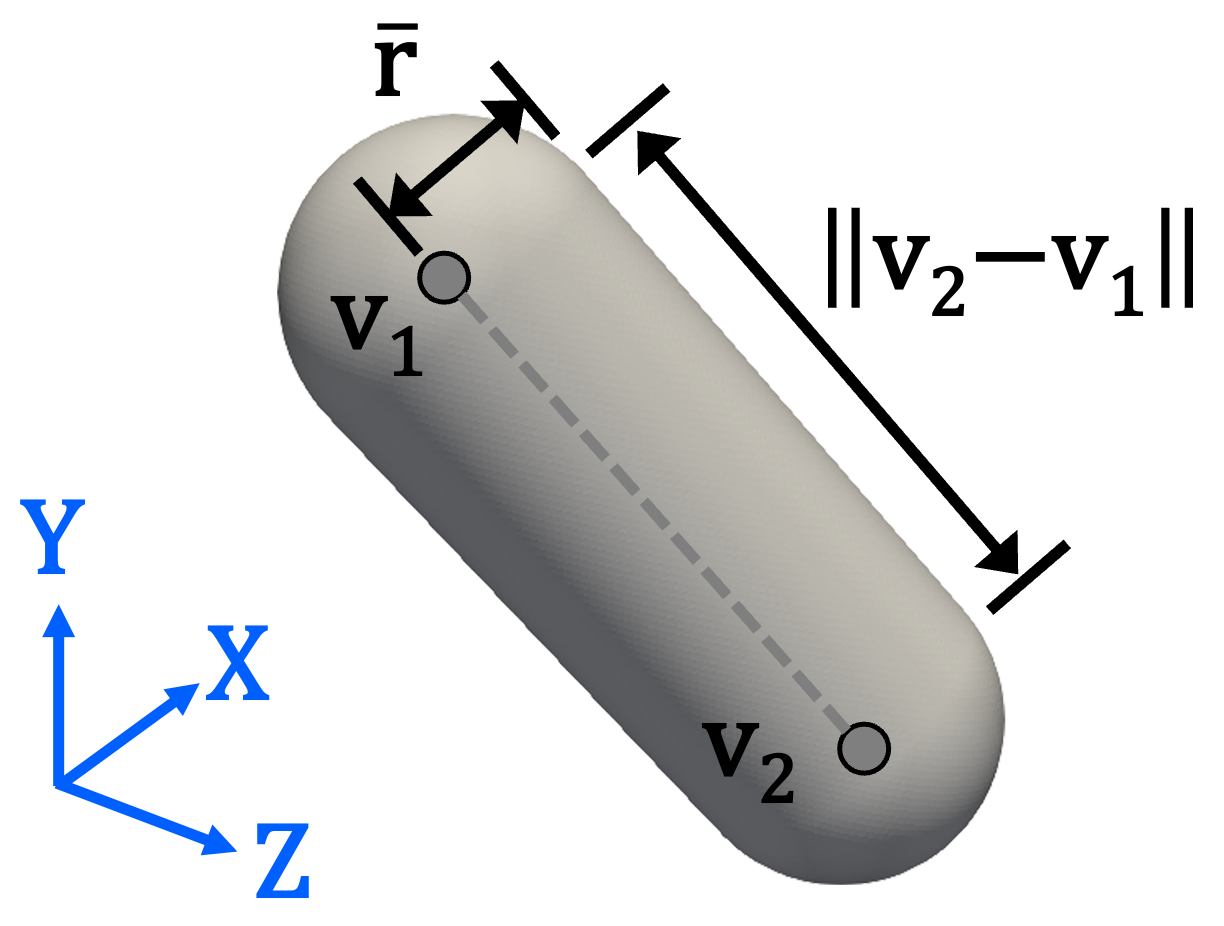}}  
  \caption{The definition of beams: (a) A beam radius is defined from its adjacent seeds; (b) A single beam description.}
  \label{fig:beam}
\end{figure}

Each Voronoi foam $\Omega^V(\bX,\br)$ is composed of two parts: the inner part and the boundary part.  For the inner part, we first compute the Voronoi diagram from the seeds $\bX=\{\bX_i,\ i=1,\ldots, N_s\}$ in a large enough bounding box, then clip all its edges using the outer free-form shape $\Omega$. $j$-th remaining edge is turned into a beam with the radius $\bar{r}_j$ averaged from the neighboring seeds:
\eb
\bar{r}_j = \sum_{i\in \calX_j}r_i/|\calX_j|,
\ee
where $\calX_j$ is the set of neighboring seeds of edge $j$, $|\calX_j|$ is the number of seeds. The radii on seeds are collected in a vector $\br$. The image in Fig.~\ref{fig:beam}(a) indicates the situation.

Corresponding to a Voronoi edge of vertices $\bv_1,\bv_2$, a beam $\phi$ is defined which consists of a cylinder with radius $\bar{r}$ and height $\|\bv_2 - \bv_1\|$ and two half-sphere ends with radius $\bar{r}$; see Fig.~\ref{fig:beam}. The implicit form $\tilde{\phi}$ of the beam $\phi$ is defined as follows, 
\eb\label{eq-levelset-singlebar}
\tilde{\phi}(\bx) =\tilde{\phi}(\bx,\bv_1,\bv_2,\bar{r}) = \bar{r} - d(\bx,\bv_1,\bv_2) ,
\ee
where $d(\bx,\bv_1,\bv_2)$ represents the minimum distance from the point $\bx$ to the edge $\bv_1,\bv_2$,
\eb
d(\bx,\bv_1,\bv_2) =
\left\{
\begin{array}{ll}
\|\bb\|, &\mbox{if}\ \ba\cdot\bb\leq 0, \\
\|\bg\|, &\mbox{if}\ 0<\ba\cdot\bb<\ba\cdot\ba,\\
\|\be\|, &\mbox{if}\ \ba\cdot\bb\geq \ba\cdot\ba,
\end{array}
\right.
\ee
for 
\eb
\ba = \bv_2 - \bv_1,\  \bb =\bx - \bv_1, \ \be  =\bx - \bv_2, \ 
\bg =(\bI - \frac{1}{\|\ba\|^2}\ba\otimes\ba)\bb.
\ee 

\begin{figure}[!htb]
  \centering
%
\subfloat[max]{\includegraphics[width=0.3\linewidth]{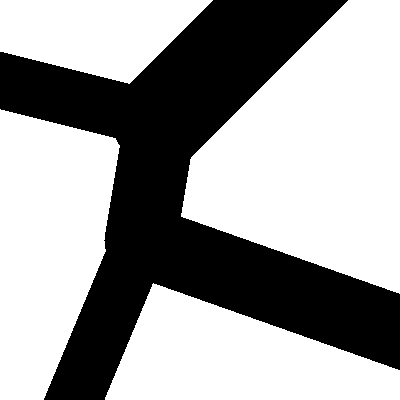}}\quad 
\subfloat[KS]{\includegraphics[width=0.3\linewidth]{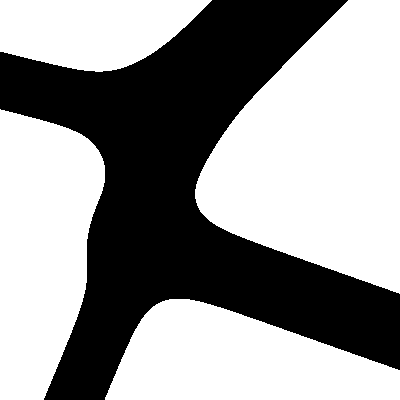}}\quad 
\subfloat[difference]{\includegraphics[width=0.3\linewidth]{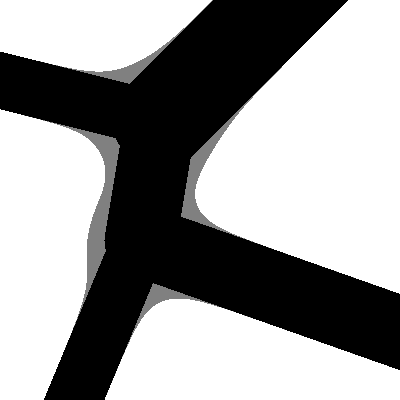}}
\caption{Comparison between direct boolean
and implicit beam blending.}
 \label{pics:blendingExamples}
\end{figure}

The implicit representation of the whole foam is the union of the implicit functions of all beams after smoothing  
%
%
using Kreisselmeier-Steinhauser (KS) function~\cite{zhu2020Areview_40}, 
\eb\label{eq-ks}
\Phi(\bx)=\frac{1}{p}\ln\left(\sum_{j=1}^n e^{p\cdot (\phi_j(\bx)-\phi_{\max}) }\right) + \phi_{\max},
\ee  
where $\phi_{\max}=\max(\phi_1,\ldots, \phi_n)$ for the beam number $n$, and $p=16$ in this paper. KS function makes the description function $\Phi(\bx)$ compact and differentiable w.r.t. $\bx$. The different structures constructed using the union function and the KS function are compared in Fig.~\ref{pics:blendingExamples}. 

The boundary part related to the outer shape $\Omega$ is also represented 
in an implicit form $\phi_{\Omega}(\bx)$. When a closed shell is required, Eq.~\eqref{eq-ks} is directly extensible by taking $\phi_{\Omega}(\bx)$ as an additional beam. Otherwise, we include the intersections the outer shapes and the Voronoi faces, where each face thickness is the average of the radii of the two seeds that determine the face. For simplicity, we will not distinguish the Voronoi foams with or without a shell. 


The \emph{Voronoi foam} $\Omega^V(\bX,\br)$ is ultimately described as a density field by a regularization Heaviside function $H(\Phi(\bx))$, 
\eb\label{eq:H}
\small
    H(\Phi(\bx))= \begin{cases}
        1,  & \text { if } \Phi(\bx) > \epsilon, \\
        {\frac{3(1-\alpha)}{4}}\left(\frac{\Phi}{\epsilon}-\frac{\Phi^{3}}{3 \epsilon^{3}}\right)+\frac{(1+\alpha)}{2}, & \text { if }-\epsilon \leq \Phi(\bx) \leq \epsilon, \\
        \alpha, & \text { otherwise },\end{cases}
\ee
where $\epsilon$ controls the magnitude of regularization, $\alpha=1e^{-6}$ by default is a small positive number to avoid a singular global stiffness matrix~\cite{zhang2016new}. 
%

\section{Designing goal under Voronoi foams}
\label{sec:goal} 
The Voronoi design problem takes $\bX,\br$ as design variables for performance optimization. Its analytical and discrete formulations are explained below. We focus on the most popular problem of stiffness optimization~\cite{Wu2021Review}.  

\subsection{Problem formulation}
Given seeds $\bX$ and radii $\br$, the material volume $V(\bX,\br)$ of the Voronoi foam is
\begin{align} \label{eq:Ve}
V(\bX,\br) & =\int_{\Omega} H(\Phi(\bx))~dV.
\end{align}
The {admissible space} that constrains the volume of the Voronoi foam, denoted $A$, is defined as  
\begin{equation}
  A = \left\{(\bX, \br) \mid V(\bX, \br) / V_{0} \leq v,\ \  
  \underline{\bX} \leq \bX \leq \overline{\bX},
  \underline{\br} \leq \br \leq \overline{\br}
  \right\},
\end{equation}
where $V_{0}$ is the volume of design domain and $v$ is a prescribed
volume fraction.  The lower and upper bounds of $\bX$,
i.e. $\underline{\bX},\overline{\bX}$, are set according to the axis-aligned bounding box of $\Omega$ or from user prescription. The bounds of $\br$, for simulation accuracy, are set so that the width of a beam at least spans two fine mesh elements, i.e.
\eb
\underline{\br} \geq 2~l_{a},
\ee
where $l_{a}$ is the average edge length of fine mesh.

The domain containing the Voronoi foam is equipped with a non-uniform
material by mapping the density $H(\Phi(\bx))$ to the fourth-order elastic tensor $\bD(\bx)$:
\begin{equation}
\bD(\bx,\bX,\br)=H(\Phi(\bx))\bD_0,
\end{equation}
where $\bD_0$ is the constant elastic tensor in the solid regions. $\bD(\bx)$ is correspondingly nearly zero in the void regions not covered by the beams.  

Then, for a user specified load $\btau$, its associated static
displacement $\bu$ under the test function $\bv$ in Sobolev vector
space $H_{0}^{1}(\Omega)$ is characterized by an equation involving
the strain vectors $[L\bu]$, $[L\bv]$:
\begin{equation}
  a(\bu,\bv, \bD) = l(\bv),\ \forall \bv\in H_0^1(\Omega),
  \label{eq:load2disp}
\end{equation}
where
\eb
a(\bu,\bv,\bD)=\int_{\Omega} [L\bu]^T~\bD(\bx,\bX, \br)~[L\bv]~\rd V,
\ee
and 
\eb
l(\bv)=\int_{\Gamma_N} \btau\cdot \bv~\rd S.
\ee

As the goal, we hope the overall deformation of the Voronoi foam
$\Omega^V(\bX, \br)$ is small. Therefore, we introduce two terms about
compliance and shape. The compliance $C(\bX,
\br, \bu)$ measuring the elastic potential of the body, as widely
adopted in topology optimization, is set as the physical objective,
\begin{equation}\label{eq:C}
    C(\bX, \br, \bu) = \frac{1}{2} \int_{\Omega}  [L\bu]^T \bD(\bx, \bX, \br) [L\bu] ~\rd V.
\end{equation}

The shape regulation energy $S(\bx)$ is to regularize the Voronoi cells to approximate regular polyhedrons
\eb\label{eq:Es}
S(\bX) = \sum_{i = 1} ^ {N_s} w_{\bX_i}~|| \bX_i - \bX_i^c ||^2,
\ee
that is, the sum of the Euclidean distances between seed $\bX_i$ and the centroid $\bX_i^c$ of each Voronoi cell $\calV_i$ weighted by $w_{\bX_i}$~\cite{du1999centroidal}. $w_{\bX_i}=1$ is simply adopted here.

Finally, we get the constrained optimization problem:
\begin{equation}
\begin{aligned}\label{eq:prob}
    \min_{(\bX, \br)\in A,\ \bu} ~ &J(\bX, \br, \bu) \\
    \text{s.t.}\quad &a(\bu,\bv, \bD) = l(\bv),\ \forall \bv\in H_0^1(\Omega),\\
 \end{aligned}
\end{equation}
where the design target $J(\bX, \br, \bu)$ is set as the weighted sum of the physical objective and the shape regularization term,
\begin{equation}\label{eq:target}
    J(\bX, \br, \bu) = (1-w)~C(\bX, \br, \bu) + w~S(\bX).
\end{equation} 
Notice here that different measures of physical performances or shape regularizations can be introduced for different design purposes, and the target foams can be derived similarly following the procedure described below.   

\subsection{Discretization}
\label{sec:discretization}
To discretize the displacement, strain or stress fields, one can of
course directly tessellate the foam by a boundary conforming fine mesh. However, we build a fixed background (linear tetrahedral) mesh $D = \{D_{e}, e=1,2,\cdots,N\}$ to cover $\Omega$, 
i.e. $\Omega \subset D$. 
This choice brings two merits: avoiding the very time-consuming and error-prone
generation of boundary conforming meshes and ensuring convergence of
optimization by simulation on a fixed mesh.

In each tetrahedral element $D_e$, the density $H_e$ is set to the average of $H(\bx)$ on its four nodes,
\eb\label{eq:rhoe}
H_e(\bX,\br) = \sum_{k=1}^4 H(\Phi(\bv^k_e,\bX,\br)) / 4,
\ee
for $\bv^k_e$ being the coordinate of $k$-th node in element $D_e$.

Giving the vector of discrete displacements $\bQ$ on the nodes of background mesh $D$, we collect the displacements on the nodes of $e$-th element $D_e$ as $\bQ_e$, then the displacement on any point of $\bx \in D_e$ can be interpolated as
\begin{equation} \label{eq:NQ}
    \bu(\bx) = \bN_{e}(\bx) \bQ_{e},
\end{equation}
where $\bN_{e}(\bx)$ denotes the element linear bases (shape functions) on the nodes of $e$-th element. The element stiffness matrix $\bK_e$ can be further derived following a classical Galerkin FE method,
\eb
\bK_e(\bX,\br) = H_e(\bX,\br)~\int_{D_e}[L\bN]^T~\bD_0~[L\bN]~\rd V. \label{eq:Ke}
\ee

The global stiffness matrix $\bK(\bX,\br)$ can be simply assembled by summing $\bK_e(\bX,\br)$. Now, the equilibrium equation Eq.~\eqref{eq:load2disp} is discretized into the following equation
\begin{equation}
  \bK(\bX,\br)\bQ=\bff,
  \label{eq:solve_Q}
\end{equation}
where $\bff$ is discretized load $\btau$. Solving above equation to get the displacement vector $\bQ$, the compliance $C$ is computed from 
\begin{equation}
  C(\bX, \br, \bu)
  \approx \frac{1}{2}\|\bQ\|_{\bK(\bX,\br)}^2=C(\bX, \br, \bQ).
  \label{eq:potential}
\end{equation}

\section{Optimization method}\label{sec:opt} After discretizing Eq.~\eqref{eq:prob}, we get the optimization problem:
\begin{equation}
  \begin{gathered}
    \min_{(\bX,\br) \in A,\ \bQ} (1-w)C(\bX,\br,\bQ)+wS(\bX)\\
    \text{s.t.}\quad \bK(\bX,\br)\bQ=\bff.
  \end{gathered}
  \label{eq-min-problem}
\end{equation}
To eliminate the equilibrium state constraint, we treat $\bQ$ as the
function of $(\bX,\br)$ via $\bQ(\bX,\br)=\bK^{-1}(\bX,\br)\bff$, and
view $C$ as function of $(\bX,\br)$, i.e. $C(\bX,\br)\triangleq
C(\bX,\br,\bQ(\bX,\br))$.  Now, the above problem is turned into
\begin{equation}
  \begin{gathered}
    \min_{(\bX,\br) \in A} (1-w)C(\bX,\br))+wS(\bX).\\
  \end{gathered}
  \label{eq:final_opt}
\end{equation}

Eq.~\eqref{eq:final_opt} is to be solved following a numerical gradient-based approach. 
Both the topology and geometry of the Voronoi
foams $\Omega^V(\bX,\br)$ are optimized simultaneously in this study. 
The GCMMA (Globally Convergent Method of Moving Asymptotes)~\cite{Zillober1993} approach is carefully chosen. It approximates the original nonconvex problem through a set of convex sub-problems by using the gradients of
the optimization objective and constraints with respect to the design variables $\bX$ and $\br$ derived below. 

According to the chain rule and an adjoint approach~\cite{bendsoe2003Topology}, the sensitivity of the objective function $C(\bX, \br)$ is derived as follows:
\begin{equation} \label{eq:sens_C}
  \frac{\partial C(\bX,\br)}{\partial a} = - \frac{1}{2}\bQ^{T}~\frac{\partial \bK}{\partial a}~\bQ = -\frac{1}{2}\bQ^{T}~\sum_{e} \frac{\partial \bK_{e}}{\partial a}~\bQ.
\end{equation}
According to Eq.~\eqref{eq:Ke}, we have
\begin{equation}
  \frac{\partial \bK_e(\bX,\br)}{\partial a} = \frac{\partial H_e(\bX,\br)}{\partial a}~\int_{D_e} [L\bN]^T~\bD_0~[L\bN]~\rd V,
  \label{eq:pKpa}
\end{equation}
and
\begin{equation}
  \frac{\partial H_e(\bX,\br)}{\partial a} = \frac{1}{4}~\sum_{k=1}^4 ~\frac{\partial H(\Phi(\bv^k_e,\bX,\br))}{\partial a},
  \label{eq:pHpa}
\end{equation}
where $a$ denotes a component of the variables $\bX_i$ or $\br_i$. 

One can easily identify two obvious challenges in the above procedure: First, the derivatives $\partial H/\partial a$ (in Eq.~\eqref{eq:pHpa}) is related to Voronoi diagram, whose derivation and expensive computations make the optimization extremely challenging. Second, complex geometry of the foam structure entails fully resolved finite element mesh, bringing about large  number of DOFs in $\bQ$ and the prohibitive cost of solving the large linear
system for $\bQ$ (in Eq.~\eqref{eq:sens_C}). 

%

In addressing the first issue, we first explore its the differentiability, and then develop an efficient numerical approach for its numerical computation by exploiting the local property of Voronoi diagram. Details are explained below. 

\subsection{Differentiablity analysis of Voronoi edge w.r.t to seeds} 
For a given parameter $a$ as a component of $\bX, \br$, the derivative $\partial H / \partial a$ boils down to terms about $\partial H/\partial \Phi$, $\partial \Phi/\partial \phi_j$ and $\partial \phi_j/\partial a$ via the chain rule from the implicit expression of $H$ in Eq.~\eqref{eq:H}.  However, it is not always valid as it implicitly assumes the topology of the underlying Voronoi diagram remains unchanged within a small variation of seed points, which is however not always true. In fact, the minimal distance $\phi_j(\bx_0,\bX,\br)$ is continuous but not always differentiable about a seed point $\bX_i$ at a specific vertex point $\bx=\bx_0$.  

\begin{figure}[!htb]
  \centering
  \includegraphics[width=0.4\textwidth]{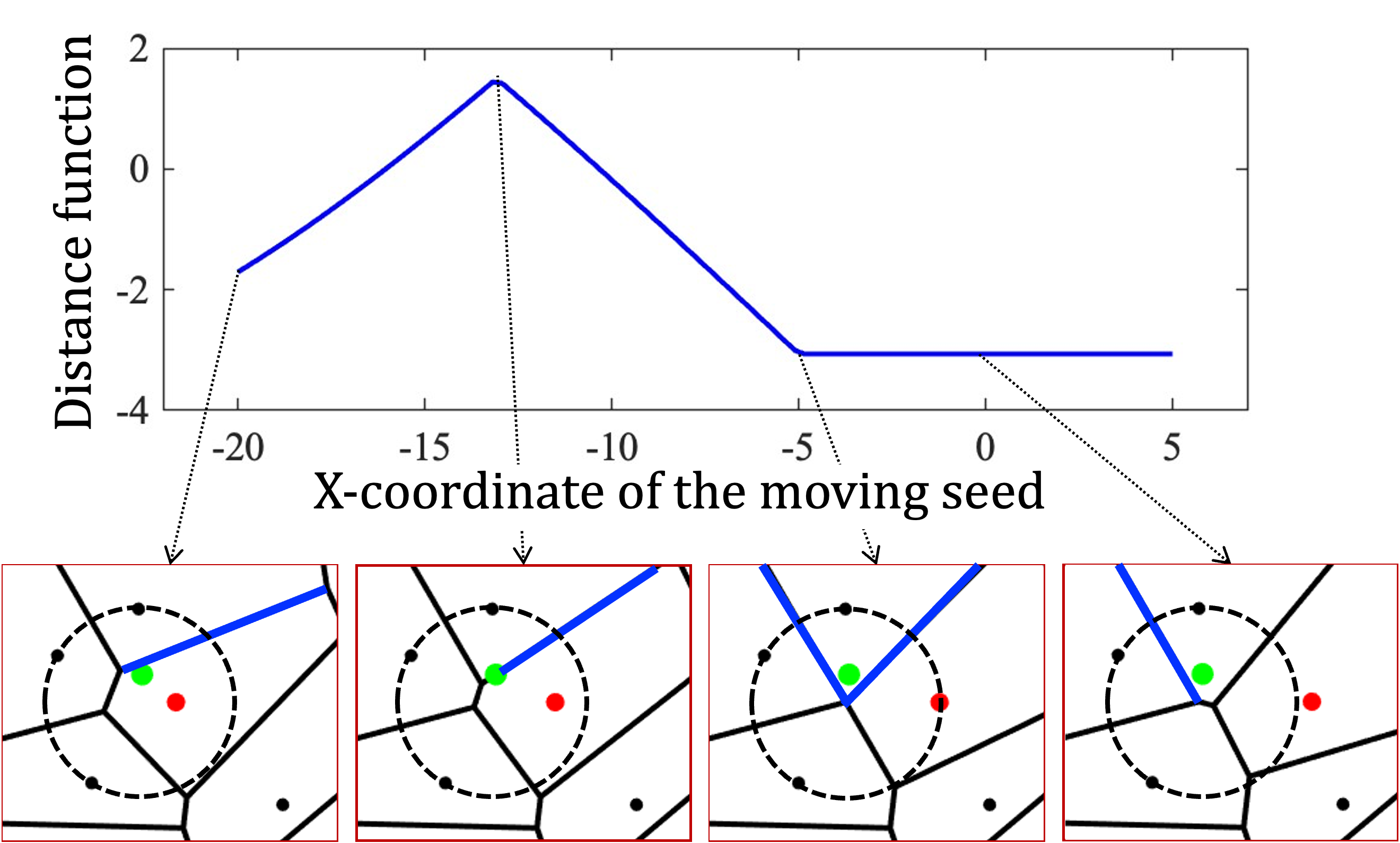}
  \caption{The distance curve from a watching point (in green) to a Voronoi foam. The third figure below shows a critical situation that the four seeds (three in black and one in red) share a circumscribed circle. The beam that $\bx_0$ is closest to is in blue.}
  \label{fig:discrv}
\end{figure}

Consider the example in Fig.~\ref{fig:discrv}, where we plot the curve $\phi(\bx_0,\bX,\br)$, a distance function from a point $\bx_0$ (in green) to the Voronoi foam $\Omega^V(\bX,\br)$, w.r.t. a seed $\bX_i$. The distance function is always continuous.
We notice the curve has some critical situations: when four seed points (three in black and one in red) share a circumscribed circle, the beam (in blue) that $\bx_0$ is closest to is jumping from one to another. 

The distance function $\phi(\bx_0,\bX,\br)$ is however not always differentiable with respect to seeds $\bX$ in two situations: the above critical situation, and when $\bx_0$ is on the Voronoi edge. Luckily, the number of such singular points is small and can be easily smoothed out~\cite{zhu2020Areview_40}. 
The result is concluded below. 

\emph{The distance function $\phi(\bx_0,\bX,\br)$ is continuous but not always differentiable w.r.t seed points $\bX_i$ at finite number of points: 1. $\phi(\bx_0,\bX,\br)$ takes its value at a critical point of $\Omega^V(\bX,\br)$ where four or more seed points share a common circumscribed circle/shpere. 2. $\bx_0$ is on the Voronoi edge.}

\subsection{Numerics in derivative computations}\label{sec-diff-num}
Computing the Jacobian matrix numerically requires great computational efforts for its huge size $(\sum_{\alpha} n^{\alpha}) \times 4N_s$ even though it is sparse, where $\sum_{\alpha}n^{\alpha}$ is the total number of fine mesh vertices with $n^{\alpha}$ being the vertex number
of the fine mesh $D^{\alpha}$, and $4N_s$ being the number of design variables. 
The number can reach as high as 1 billion ($1M \times 10K$) for the cube example in Fig.~\ref{fig:diff-v}. Further noticing that the distance function consists of a large number of beams, a direct computation for each derivative, either analytically or via finite difference, would be even prohibitive. 

The efficiency is much improved by exploring the locality of Voronoi diagram. Firstly, the \emph{2-ring criteria} of Voronoi diagram tells that only seed points in a 2-ring around $\bx_0$ influence the density on $\bx_0$~\cite{martinez2016procedural}. Accordingly, we can generate the Voronoi diagram locally by carefully picking up these local seeds. It also much reduces the number of beams $\phi_j(\bx)$ to be used in the overall distance function computation in Eq.~\eqref{eq:H}.  

\begin{figure}[!htb]
  \centering
    \includegraphics[width=0.2\textwidth]{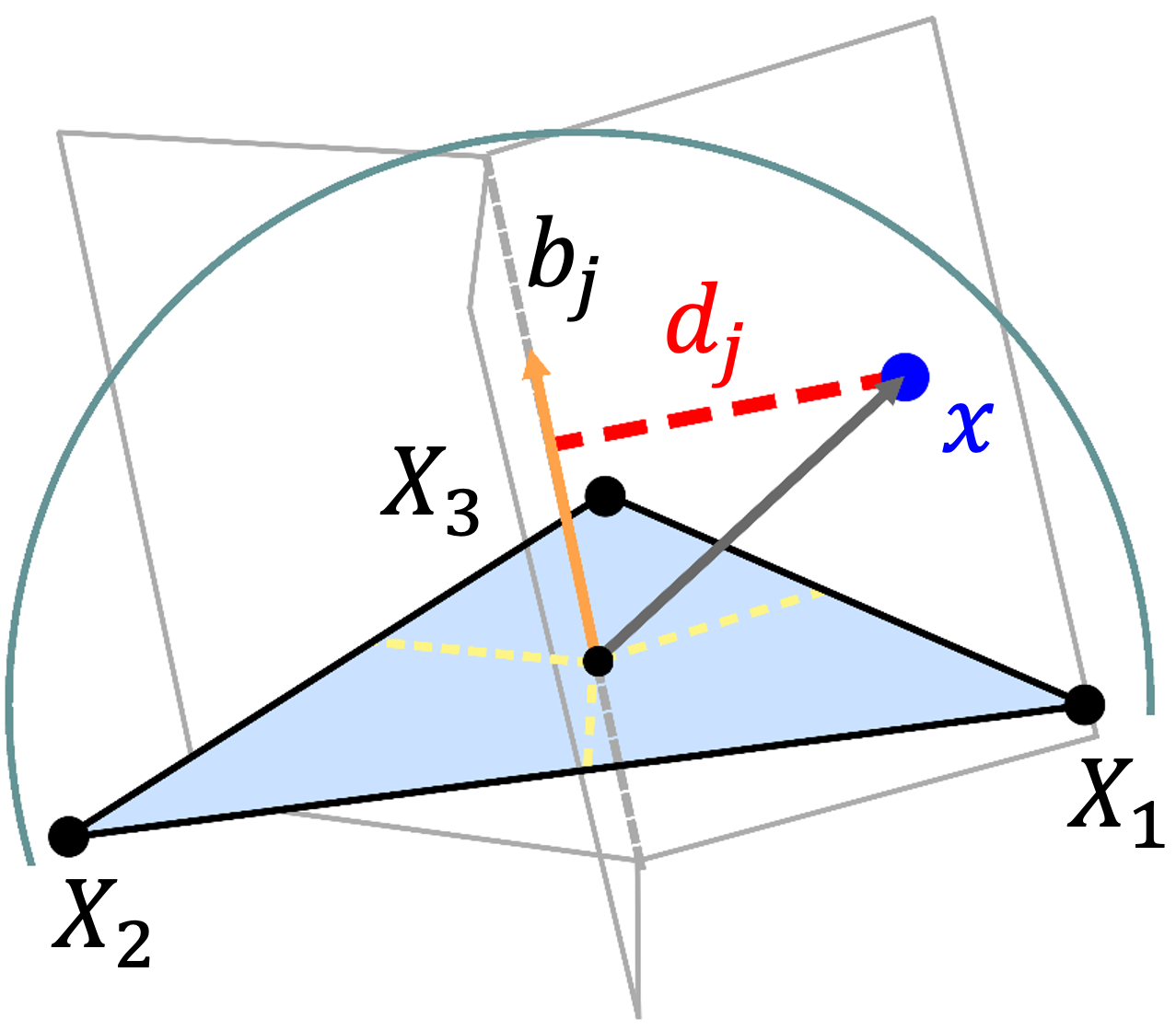} 
  \caption{Three-point distance approximation without explicit Voronoi tessellation. The signed distance is approximated as $\bar{r}_j-d_j$ for the beam radius $\bar{r}_j$.}
  \label{fig:speed-up}
\end{figure} 

We further develop a three-point distance approximation approach to improve the distance computation efficiency by only considering the three nearest points to a given vertex $\bx_0$. It is assumed that the beam radii are approximately the same. Accordingly, as indicated in Fig.~\ref{fig:speed-up}(a), let $b_j$ be the beam determined by the three seeds $\bX_1,\bX_2,\bX_3$, and $d_j$ the distance from $\bx_0$ to $b_j$. We set the value of the distance field at $\bx_0$ approximately as $\bar{r}_j-d_j$ where $\bar{r}_j$ is the averaged radii to the three points. 

It can be roughly estimated that the approximated minimal distance has a maximal error $|r_2-r_1|$ to its true value where $r_1,r_2$ are the minimal distance of the vertex $\bx_0$ to its Voronoi diagram and Voronoi foam. Accordingly, the density at $\bx_0$ will not change, or $\tilde{\rho}=\rho$, when $r_2\leq 2r_1$.

NOTICE (From Jin HUANG): Sometimes, $\phi$ is beam.  Sometimes, $\phi$ is distance to edges. Is it good to make them consistent?  Besides, I remember that $d_j$ is the distance to the line containing $b_j$ instead of the line segment $b_j$.

Algorithm~\ref{alg-derv} summarizes the above results. It computes the derivatives via finite difference for two different cases: using the three-point approximation stated above when the local radii are approximately the same or using local Voronoi reconstruction otherwise.  

\begin{algorithm} \caption{Derivative computation of density field function}\label{alg-derv}
  \KwIn{Voronoi foam $\Omega^V(\bX,\br)$, a tetrahedral vertex $\bx_0$;} 
  \KwOut{Derivative of the minimal distance function $H(\Phi(\bx_0, \bX,\br))$ w.r.t. a seed $\bX_i$;}
  (1) Select the k-nearest seeds around $\bx_0$ and the associated distance functions that may influence the density value at point $\bx_0$ based on the 2-ring criteria. Collect the influencing distance functions as $\phi_j(\bx,\bX,\br), \ j\in J$.  \\ 
  
  (2) Compute the derivative of $\phi_j(\bx_0,\bX,\br),\ j\in J$ w.r.t. $\bX_i$ via finite difference by updating the density for a seed point variation as follows:  
  \begin{itemize}
  \item Let $r_1=\min(\{r_k\})$ and $r_2=\max(\{r_k\})$. If $r_2\leq 2r_1$, update the density at point $\bx_0$ via three-point approximate approach;
  \item Otherwise, update the density at point $\bx_0$ via locally reconstructing the Voronoi foam for the $k$ seeds.  
  \end{itemize}
\end{algorithm}
  

\subsection{Flowchart of the optimization algorithm}
The overall flowchart of our explicit topology optimization of a conforming Voronoi foam proceeds as Algorithm~\ref{alg-opt}, as also  illustrated in Fig.~\ref{fig:teaser}. 

\begin{algorithm}
  \caption{Explicit topology optimization of conforming Voronoi foams}\label{alg-opt}
  \KwIn{Seed number, design domain $\Omega$, background coarse mesh each with a tetrahedral fine mesh, target value fraction, and a specific linear elasticity analysis problem, as stated in Eq.~\eqref{eq:final_opt}.}
  \KwOut{The optimized Voronoi foam described by the seed locations and the associated radii $\bX,\br$.}

(1) Construct the nested background mesh $D^H$ for domain $\Omega$ (Sec.~\ref{sec-coarsening}).
  
(2) For the current seeds $\bX$ and radii $\br$, (locally) generate the Voronoi foam and the corresponding density function (Secs.~\ref{sec-space} and~\ref{sec-diff-num}).
   
(3) Construct the polyhedral shape functions $\bN^{\alpha}(\bx)$ for each coarse element $D^{\alpha}$ (see Sec.~\ref{sec-coarsening}).
  
(4) Compute the displacement $\bQ$ based on the polyhedral shape functions $\bN^{\alpha}(\bx)$ (Sec.~\ref{sec-coarsening}).
  
(5) Compute the derivatives of design target and constraints w.r.t $\bX,\br$ (Secs.~\ref{sec-diff-num} and \ref{sec-coarsening}), and update their values via GCMMA.
  
(6) Repeat steps 2-5 until convergence.
\end{algorithm}

\section{Numerical coarsening}\label{sec-coarsening}
As in Section~\ref{sec:discretization}, the conventional FEM evaluates the element stiffness matrix $\bK_e$ on fine tetrahedral elements, and assembles them into the global stiffness matrix $\bK$. However, in the numerical coarsening method proposed in this section, we use the counterparts $\bK^{\alpha}$ on coarse elements instead of $\bK_e$ to constitute $\bK$. There are many numerical coarsening methods, and one of the state-of-the art methods~\cite{li2022analysis}, which shows advantages over previous approaches, is taken here. This method also takes the strategy of material aware bases~\cite{chen2018numerical}.

\begin{figure}[!htb]
    \centering
    \includegraphics[width=0.3\textwidth]{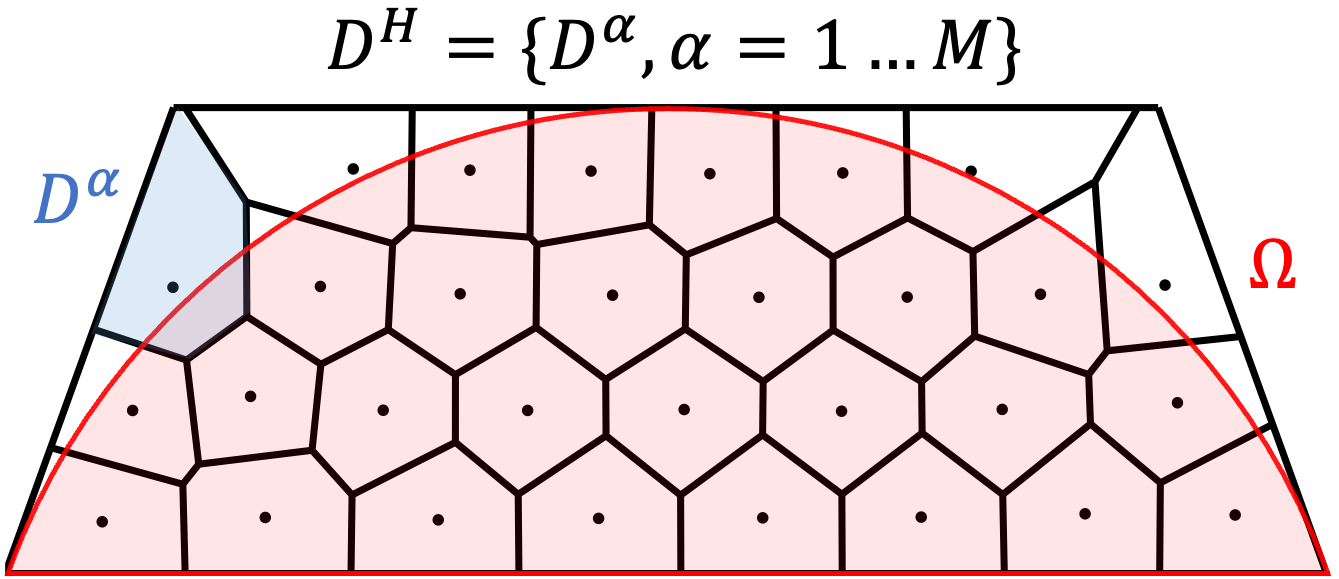}
    \caption{Given a domain $\Omega$, a background mesh $D^H$ composed of coarse elements $D^{\alpha}$ is generated for numerical coarsening. $D^H$ contains $\Omega$.}
    \label{fig:mesh-def}
\end{figure}

\begin{figure}[!htb]
    \centering
    \includegraphics[width=0.35\textwidth]{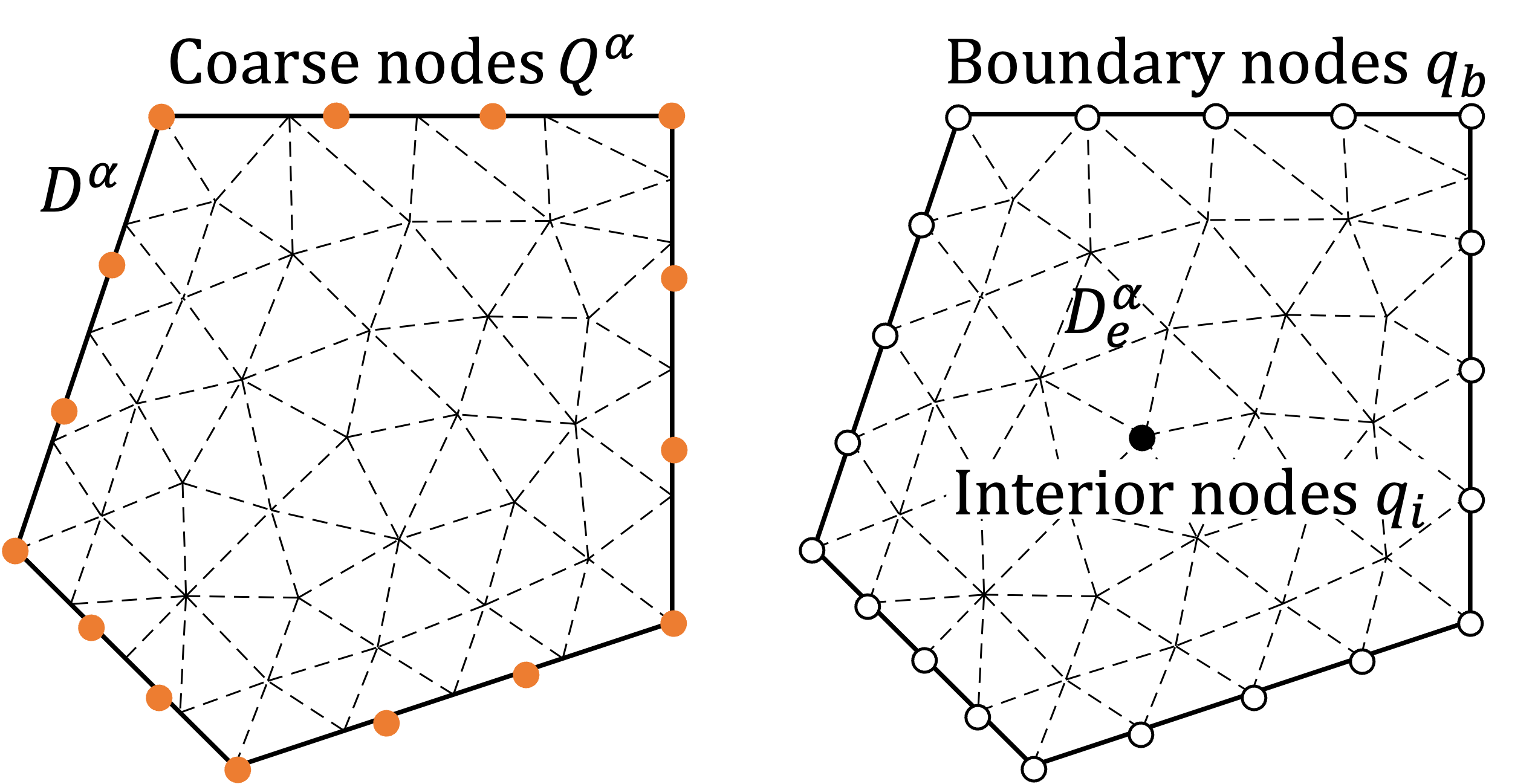}
    \caption{Node definition. Boundary nodes and interior nodes are defined on the fine elements ${D_e^\alpha}$, and coarse nodes are defined along the boundaries of the coarse element $D^{\alpha}$. The coarse nodes do not have to be coincident with boundary nodes.}
    \label{fig:nodes-def}
\end{figure}

Two kinds of meshes are involved: (1) the coarse polyhedral mesh $D^H=\{D^{\alpha}, \alpha=1,2,\cdots,M\}$, (2) the refined  tetrahedral mesh $D^{\alpha} = \{D^{\alpha}_{e}, e=1,2,\cdots,e^{\alpha}\}$ engulfed by each coarse polyhedral element. Figs.~\ref{fig:mesh-def} and ~\ref{fig:nodes-def} present the coarse mesh $D^H$ and fine mesh $D^{\alpha}$ for a 2D case. 
The number of fine elements $N=\sum_{\alpha=1}^{M} e^{\alpha}$ is much greater than that of coarse elements here, i.e. $M \ll N$. Two kinds of nodes are also involved, namely (1) nodes defined along boundaries of coarse elements, abbreviated as coarse nodes, with their displacements $\bQ^H$, (2) boundary nodes and interior nodes of fine tetrahedral mesh, with their displacements $\bq_b$ and $\bq_i$, as indicated in Fig.~\ref{fig:nodes-def}.

\begin{figure*}[!htb]
  \centering
  \includegraphics[width=0.8\textwidth]{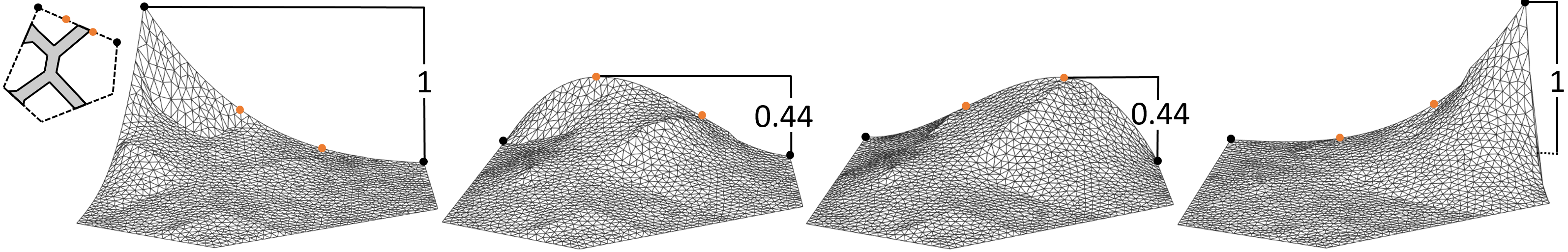}
  \caption{Surfaces of the shape function component $\bPhi_{11}(\bx)$, on four coarse nodes along one polyhedral edge, including the corner nodes (black) and the additional control nodes (orange). All the surfaces clearly exhibit flatter variations above the area with stiffer beams, while drop rapidly over the void area (filled with the extra-soft material to avoid numerical singularity), reflecting structure deformation and its material-awareness. 
  }
  \label{fig:NH-surf}
\end{figure*}

In our numerical coarsening method, coarse polyhedral mesh is used for simulation. For retaining sufficient accuracy, each coarse node is associated with a basis function (shape function) aware of the material inside, as indicated in Fig.~\ref{fig:NH-surf}. Significant efficiency improvement can be achieved owing to the smaller amount of DOFs.

Specifically, given the vector of discrete displacements $\bQ^H$ on coarse nodes, we collect the displacements on coarse nodes of $\alpha$-th coarse element $D^{\alpha}$ as $\bQ^{\alpha}$.  The displacements on a point $\bx \in D^{\alpha}$ can be interpolated as
\eb\label{eq-uNQ}
\bu(\bx) = \bPhi^{\alpha}(\bx)~\bQ^{\alpha},
\ee
where $\bPhi^{\alpha}(\bx)$ is the material-aware element basis function we use, defined on coarse nodes of $D^{\alpha}$. 

$\bPhi^{\alpha}(\bx)$ is essentially a linear composition of linear bases on the fine mesh:
\eb\label{eq-gshapef}
\bPhi^{\alpha}(\bx) = \bN^{\alpha}(\bx)~\bPsi^{\alpha},
\ee
where $\bN^{\alpha}(\bx)$ denotes the assembly of linear bases on the fine mesh of $D^{\alpha}$, and $\bPsi^{\alpha}$ is the transformation matrix from displacements $\bQ^{\alpha}$ of coarse nodes to those $\bq^{\alpha}$ of fine nodes,
\eb \label{eq:bQ2bq}
\bq^{\alpha} = \bPsi^{\alpha}~\bQ^{\alpha}.
\ee

Substituting Eq.~\eqref{eq-gshapef} and Eq.~\eqref{eq:bQ2bq} into Eq.~\eqref{eq-uNQ}, we have
\eb
\bu(\bx) = \bPhi^{\alpha}(\bx)~\bQ^{\alpha} = \bN^{\alpha}(\bx)~\bq^{\alpha},
\ee
i.e. the interpolated displacements are essentially obtained by linear shape functions on fine mesh. 

Thus, the coarse element $D^{\alpha}$ stiffness matrix $\bK^{\alpha}$ is given by Eq~\eqref{eq:Ke} using $\bPhi^{\alpha}$ instead of $\bN^{\alpha}$, i.e.
{\small
\eb \label{eq:Ke^alpha}
\begin{split}
& \bK^{\alpha}(\bX,\br) \\
&= \int_{D^{\alpha}}[L\bPhi^{\alpha}]^T~\bD(\bx,\bX,\br)~[L\bPhi^{\alpha}]~\rd V \\
&= \sum_{e}^{e^{\alpha}} H_e(\bX,\br)~\int_{D^{\alpha}_e}(\bPsi^{\alpha})^T~[L\bN^{\alpha}]^T~\bD_0~[L\bN^{\alpha}]~\bPsi^{\alpha}~\rd V \\
&= (\bPsi^{\alpha})^T~\left(\sum_{e}^{e^{\alpha}} H_e(\bX,\br)~\int_{D^{\alpha}_e}[L\bN^{\alpha}]^T~\bD_0~[L\bN^{\alpha}]~\rd V\right)~\bPsi^{\alpha} \\
&= (\bPsi^{\alpha})^T~\bk^{\alpha}~\bPsi^{\alpha},
\end{split}
\ee
where $\bk^{\alpha}$ is the high fidelity stiffness matrix for the fine mesh of $D^{\alpha}$. In like wise, the gradient of element stiffness matrix in Eq.~\eqref{eq:pKpa} is computed in coarsened simulation as
{\small
\eb
\begin{split}
& \frac{\partial \bK^{\alpha}(\bX,\br)}{\partial a}\\
&=(\bPsi^{\alpha})^T~\left(\sum_{e}^{e^{\alpha}} \frac{\partial H_e(\bX,\br)}{\partial a}~\int_{D^{\alpha}_e}[L\bN^{\alpha}]^T~\bD_0~[L\bN^{\alpha}]~\rd V\right)~\bPsi^{\alpha} \\
&= (\bPsi^{\alpha})^T~\frac{\partial \bk^{\alpha}}{\partial a}~\bPsi^{\alpha}.
\end{split}
\ee
}
}

\subsection{Shape functions as node value mapping} 
The remaining critical issue is the construction of transformation matrix $\bPsi^{\alpha}$, which takes into account the material inside. The material distribution changes with $(\bX,\br)$, so $\bPhi^{\alpha}$ (i.e. $\bPsi^{\alpha}$) needs to be updated accordingly. Unlike~\cite{chen2018numerical}, \cite{li2022analysis} does not require solving global harmonics on the fine mesh, so it is much faster and adopted here. However, the voxel coarse mesh with curved bridge nodes (CBNs) as coarse nodes is adopted in \cite{li2022analysis} for regular shapes, which cannot tightly approximate free-form domain. For higher simulation accuracy, we extend the approach to handle more general coarse elements (e.g. tetrahedrons, or more versatile polyhedrons).

In our approach, $\bPsi^{\alpha}$ in Eq.~\eqref{eq-gshapef} is derived as a product of \emph{boundary--interior transformation matrix} ${\bM}^{\alpha}$ and \emph{boundary interpolation matrix} $\bpsi^{\alpha}$, as,
\eb\label{eq:NH}
{\bPsi}^{\alpha} = {\bM}^{\alpha} ~ \bpsi^{\alpha}, 
\ee
where $\bpsi^{\alpha}$ and ${\bM}^{\alpha}$ maps the displacements from the coarse nodes $\bQ^{\alpha}$ to the boundary nodes $\bq_b$ and then to the full fine nodes $\bq^{\alpha}$. Construction of the two transformation matrices is explained below. 


\subsection{Boundary--interior transformation matrix} \label{sec-boundary_to_interior}
Firstly, ${\bM}^{\alpha}$ is derived from the local FE analysis on the fine mesh of $D^{\alpha}$ just following procedures in Section~\ref{sec:discretization}, with the equilibrium equation 
\begin{align}\label{eq:micros}
  \begin{bmatrix}
    \bk_{b} & \bk_{bi} \\
    \bk_{ib} & \bk_{i}
  \end{bmatrix}\begin{bmatrix}
    \bq_b \\
    \bq_i
  \end{bmatrix} = \begin{bmatrix}
    \bff_b \\
    0
  \end{bmatrix},
\end{align}
where $\bk_{b}, \bk_{i},\ \bk_{bi},\ \bk_{ib}$ are the sub-matrices of the fine tetrahedral mesh stiffness matrix $\bk^{\alpha}$, and $\bff_b$ the vector of exposed forces on the boundary nodes.

We have the relation of $\bq_i=  ({-\bk_{i}^{-1}}\bk_{ib}) \bq_b$ from the second-row of Eq.~\eqref{eq:micros}. Accordingly, we have the transformation from displacements $\bq_b$ of the boundary nodes to those $\bq^{\alpha}$ of the full fine nodes,
\eb
\bq^{\alpha} = 
\begin{bmatrix}
    \bq_b \\
    \bq_i
\end{bmatrix} =
\bM^{\alpha} \bq_b,
\ee
and resulted \emph{boundary--interior transformation matrix} $\bM^{\alpha}$ has the form
\eb\label{eq:bitrans}
{\bM^{\alpha}}=
\begin{bmatrix}
    {\bI_b} \\
    {-\bk_{i}^{-1}}\bk_{ib}
  \end{bmatrix},
\ee
where $\bI_b$ is the $b \times b$ identity matrix. 

\subsection{Boundary interpolation matrix} \label{sec-boundary_interpolation}
The \emph{boundary interpolation matrix} $\bpsi^{\alpha}$ builds interpolated displacements $\bq_b$ of all the boundary nodes from those $\bQ^{\alpha}$ of coarse nodes, i.e.
\eb
\bq_b = \bpsi^{\alpha} \bQ^{\alpha}.
\ee
It was designed for standard voxels in~\cite{li2022analysis}, and is extended for tetrahedral elements and general polyhedral elements via a generalized B\'ezier surface patch called S-patch.

\begin{figure}[!htb]
    \centering
    \includegraphics[width=0.48\textwidth]{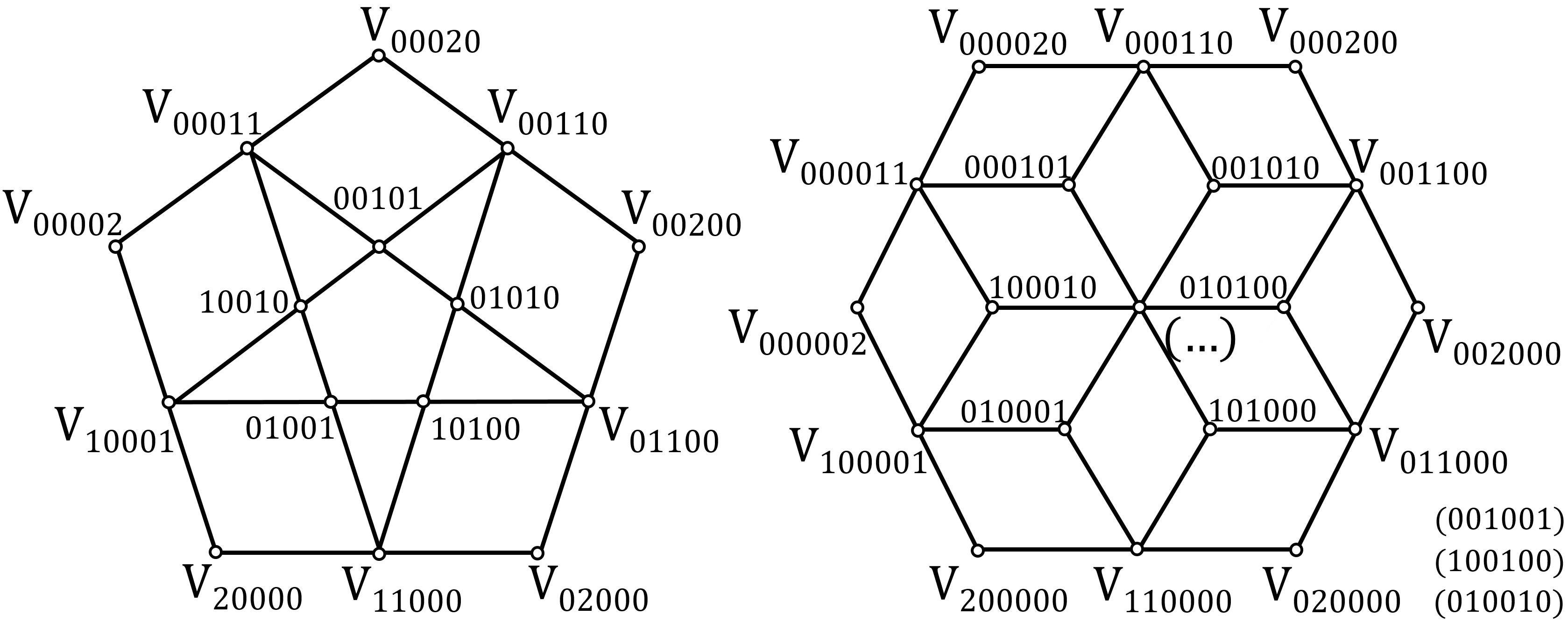}
    \caption{Labeling of control points for multi-sided S-patches of depth $d=2$, where three control points $(001001)$, (100100), and (010010) overlap (right).}
    \label{fig:S-patches}
 \end{figure}

A \emph{S-Patch} produces an interpolating multi-sided B\'ezier patch from nodal values on a polygonal face~\cite{loop1989multisided}. Given a $p$-sided polygon $P$, let $w_{k}(x),\ 1\leq k\leq p,$ be its generalized barycenter coordinate base functions~\cite{meyer2002generalized}. 
The basis function $B_{ \vec{i}}^{d}(\bx)$ is the polynomial expansion of $(\sum_{k-1}^n w_k(x))^d$ giving a set of basis functions of degree $d$,
\begin{equation}
    B_{ \vec{i}}^{d}(\bx)=\left(\begin{array}{l}
            d \\
             \vec{i}
        \end{array}\right) \prod_{k=1}^{p}\left(w_{k}(\bx)\right)^{i_{k},}, \quad | \vec{i}|=d,
\end{equation}
in which 
$\left(\begin{array}{l}
            d \\
             \vec{i}
\end{array}\right)$ is the multi-nominals expansion coefficient, index $ \vec{i} = (i_1,...,i_p)$ is a vector containing $p$ non-negative integers, $| \vec{i}|$ is the sum of indices in $ \vec{i}$. Fig.~\ref{fig:S-patches} illustrates the multi-indices for polygons with five or six edges.

Accordingly, displacement of any point $\bx$ on the S-Patch is interpolated as
\begin{equation}\label{eq:psi}
\bq(\bx) =\sum_{|\vec{i}|=d} \bQ^{\alpha}_i B_{\vec{i}}^{d}(\bx),
\end{equation}
where $\bQ^{\alpha}_i$ is the displacement of $i$-th control points on the S-patch. Evaluating displacements $\bq_b$ on all the boundary nodes of a coarse element $D^{\alpha}$ gives the \emph{boundary interpolation matrix} $\bpsi^{\alpha}$. 

Note here the fine tetrahedral meshes of adjacent coarse elements may not be identically matching along their common boundary, which much simplifies background mesh generation. This may lead to displacement discontinuities along the common boundaries. It can generally be ignored due to the high resolution of the fine meshes, and can also be improved using higher-order fine-mesh shape functions~\cite{White2023ARO}.   

\section{Results and evaluations}\label{sec-examples}
In this section, we evaluate the performance of our approach for Voronoi foam design using various examples. 

We check the optimization convergence based on the relative target change value in the latest 5 iterations~\cite{du2022efficient}. For $k$-th iteration, 
\begin{align}
    \small
    \operatorname{ch}(k)= 
    \begin{cases}
        1.0, & \text { if } k<5, \\ 
        \operatorname{ch}(k-1), &  \ \begin{tabular}[c]{@{}c@{}} if\ $k \geq 5$ \ \& \   
        $V_{e r}>1e^{-4}$\end{tabular}, \\ 
        \frac{ |\max (\bJ)-\bar{\bJ}|}{\bar{\bJ}}, & \text { otherwise },
    \end{cases}
\end{align}
where $V_{er} = (V(\bX,\br)/V_0-v)/v$, $\bJ=(J(k-4), \ldots, J(k))$, $\bar{\bJ} = |\bJ| / 5$, and  
$|\bJ|=\sum_{k-4}^{k} J(i)$ for the design target $J=(1-w)C(\bX,\br)+wS(\bX)$. Notations are also referred to Section~\ref{sec:goal}.

The error is estimated by comparing the resulted compliance $C_1$ with the reference compliance $C_0$, 
\eb\label{eq-error}
r = \frac{(C_1-C_0)^2}{ C_0^2}.
\ee

In the tests, all the femur models in Fig.~\ref{fig:teaser} have covering shells, and all the other examples do not have. We set weight $w=0.5$ in Eq.~\eqref{eq:target} for the femur model, and $w=0.1$ for the other examples. 

\begin{figure}[!htb]
  \centering
  \includegraphics[width=0.49\textwidth]{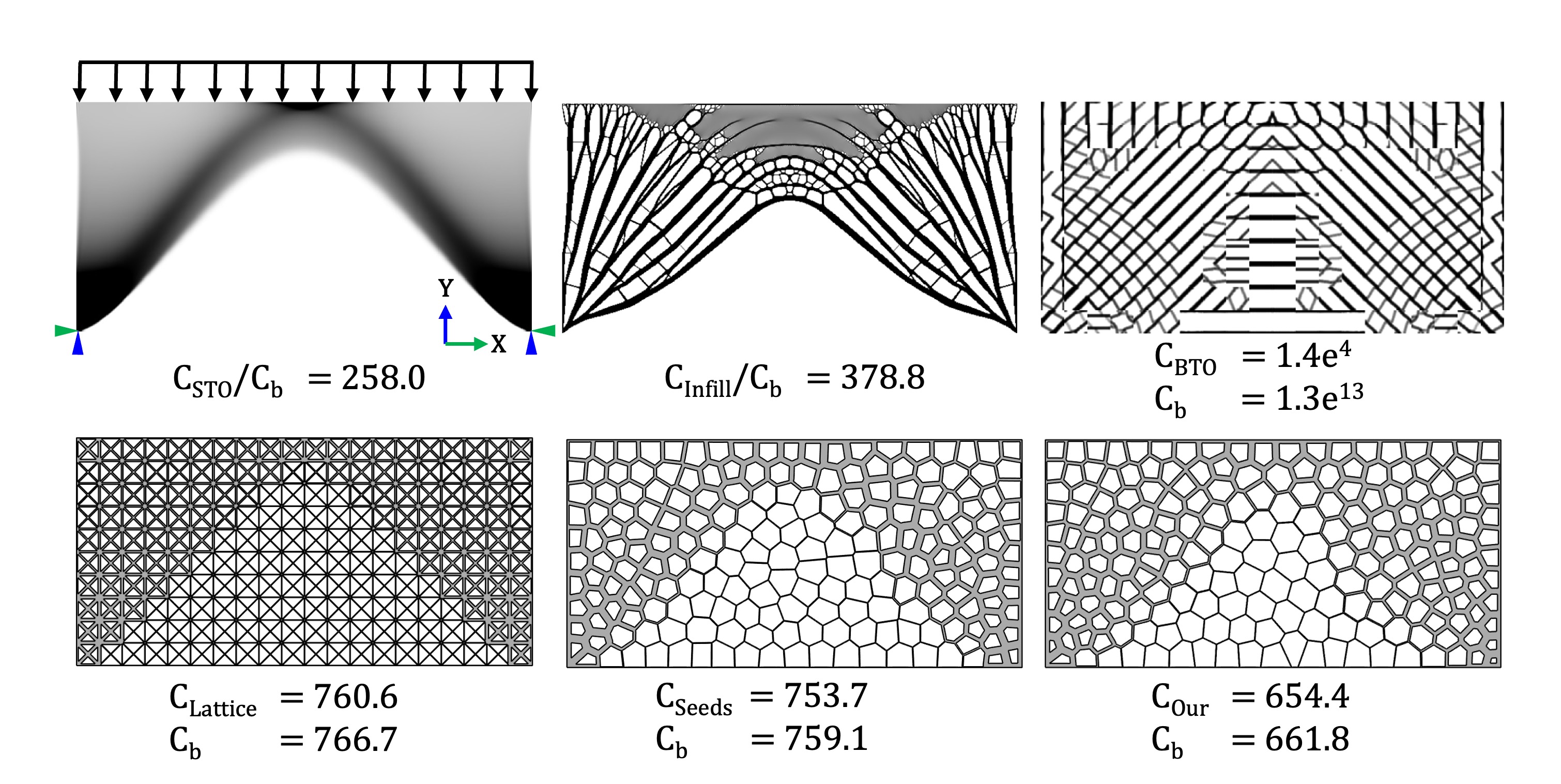}
  \caption{Comparisons between different optimization approaches.
  \emph{Infill} and \emph{BTO} failed to produce structures of valid geometry. \emph{Lattice} and \emph{Fixed-Seeds} generated foams of worst properties. Ours generated a foam of valid geometry and best properties. Here, $C$ is the compliance at convergence and $C_{b}$ is benchmark compliance computed using FEM.}
  \label{fig:compare1}
%
  \includegraphics[width=0.49\textwidth]{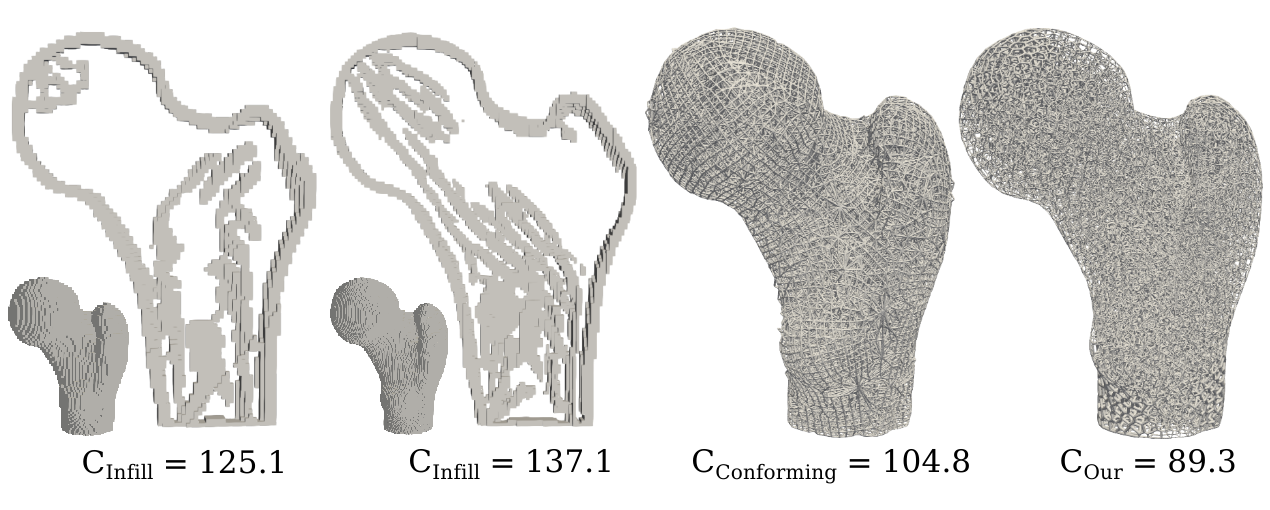}
  \caption{Comparisons with related approaches \emph{Infill} and \emph{Conforming}. Undesired broken parts were produced in \emph{Infill}. Ours produced a foam of valid geometry and best property. 
}
\label{fig:compare2}
\end{figure} 

\begin{figure}[!htb]
  \centering
  \includegraphics[width=0.45\textwidth]{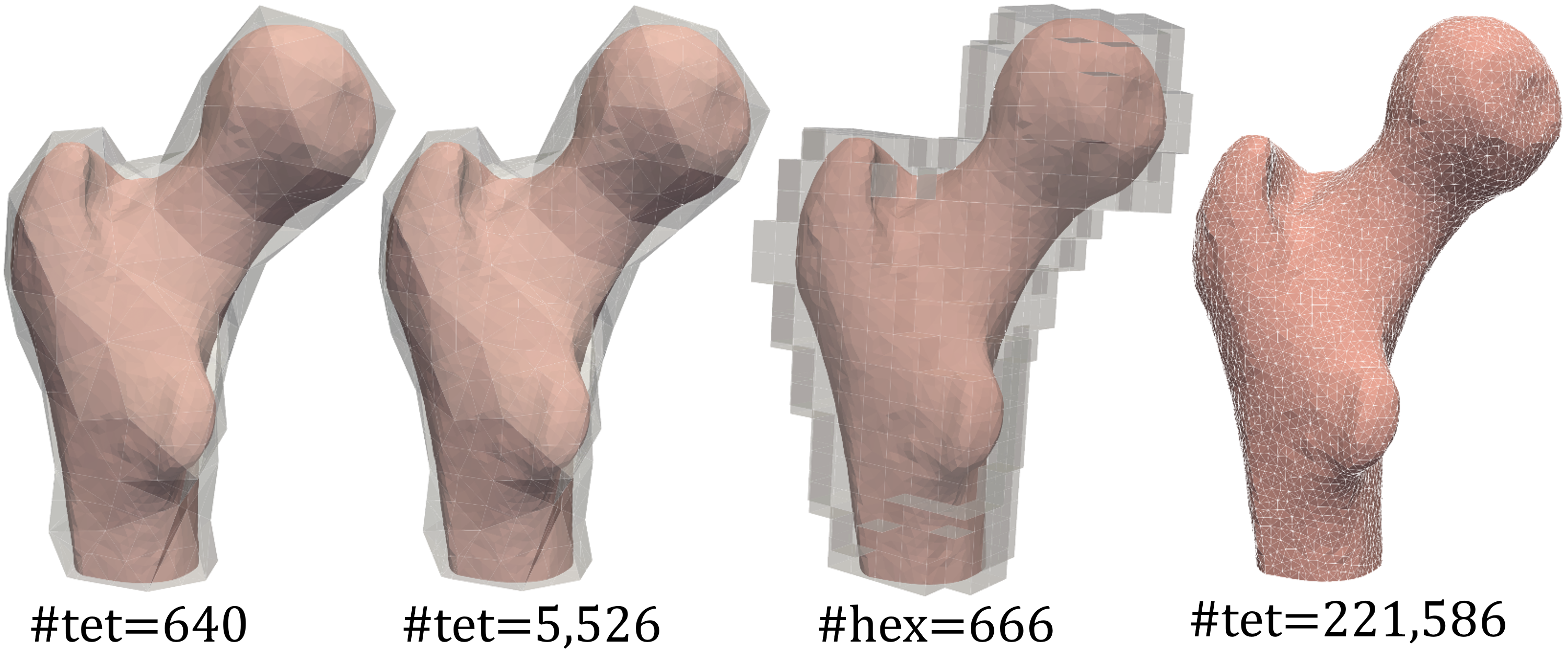}
  \caption{Different FE meshes during optimization, from left to right: our coarse mesh, a tet-mesh 8x finer than the coarsest, a hex-mesh similar to the coarsest one, the finest one (as reference) similar to our fine mesh. }
  \label{fig:scales}
\end{figure}

\begin{figure*}[!htb]
    \centering
    \includegraphics[width=0.9\textwidth]{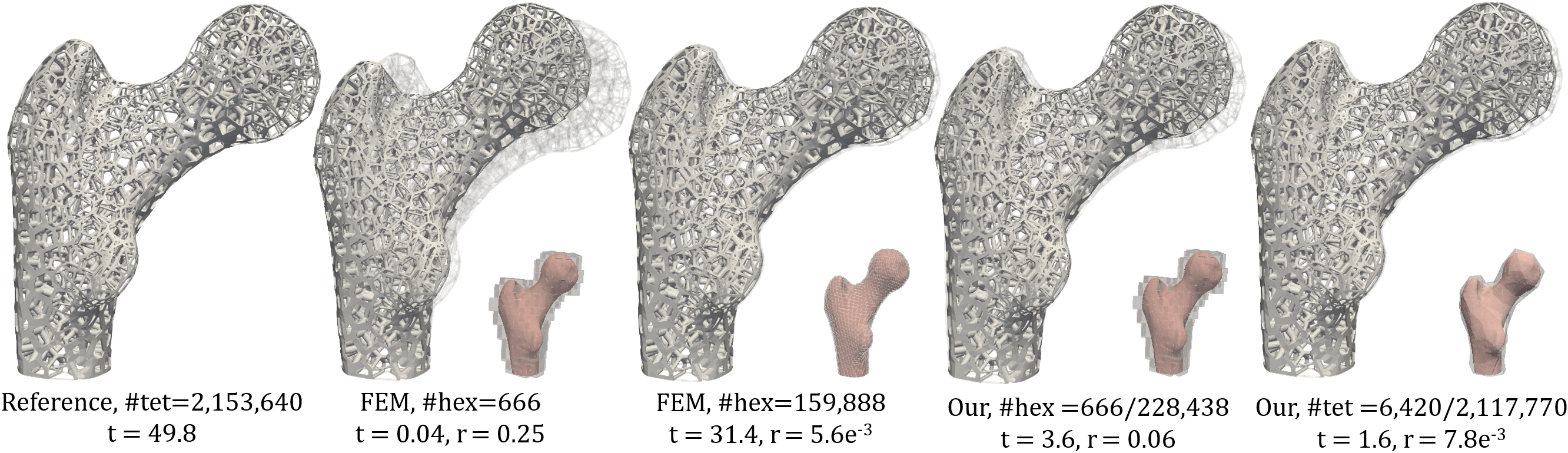}
    \caption{Accuracy test: FEM comparison between fine tet-mesh (reference), background coarse hex-mesh, background fine hex-mesh, and ours on background coarse hex-mesh and ours on background coarse tet-mesh, from left to right. $t$ is the time (seconds) of each simulation and $r$ is the error defined in Eq.~\eqref{eq-error}.}
    \label{fig:simu-mesh}
 \end{figure*}

\subsection{Comparisons with other alternatives}\label{sec-example-evf}
We evaluate the approach's performance by its comparisons with other alternatives via topology optimization, fixing cell types or different simulations. It includes: 
\emph{Infill}: SIMP-based topology optimization~\cite{Wu:2017};  
\emph{BTO}: concurrent biscale topology optimization~\cite{gao2019concurrent};  
\emph{Conforming} hexahedral mesh embedding~\cite{wu2019design});  
\emph{Lattice}: size optimization using fixed lattices~\cite{wu2019topology};  
\emph{Fixed-Seeds}: size optimization using fixed Voronoi seeds; 
\emph{STO}: SIMP-based topology optimization (by setting penalty parameter $p=1$) ~\cite{andreassen2011efficient}. 

Note that simulations in \emph{STO} and \emph{Infill} were conducted on the fine mesh, which work for arbitrary domain, while \emph{BTO} and \emph{Lattice} were conducted on the coarse mesh via numerical homogenization, which are restricted to a regular design domain. Our CBN-based approach also works for arbitrary domain; see Section~\ref{sec-coarsening}. In all the examples below, we used $C$ to denote the associated computed compliance and $C_{b}$ to denote the benchmark compliance computed using FEM on a fine mesh. 
 
\textbf{2D Comparisons.} Consider a 2D bridge problem in a regular domain in Fig.~\ref{fig:compare1}. Resolutions of the fine mesh and coarse mesh are  $320\times 960$ and $10\times 30$, and the number of seeds is $300$. In the tests, \emph{Infill} had convergence difficulty in generating large gray area, due to its large number of local volume constraints~\cite{Wang2021StressTA}. \emph{BTO} produced an invalid foam mainly due to its low-accuracy numerical homogenization, where huge difference between $C$ and $C_{b}$ was observed. Optimizations from \emph{Lattice} or \emph{Fixed-Seeds} generated geometrically valid foams but at a huge cost of increased compliance (worse property). Our method generated foam of valid geometry and of much smaller compliance (better property). 

\textbf{3D Comparisons.} Our approach was also tested for 3D models in comparison with two very related approaches \emph{Infill} and \emph{Conforming}. The femur model in ~\cite{wu2019design} was used, slightly different from ours in Fig.~\ref{fig:teaser}. The target volume fraction was set $0.5$, and the results were plotted in Fig.~\ref{fig:compare2}. 

In order to produce foams of similar bar numbers and bar sizes, the following settings were taken. Ours has $5,000$ seeds, $2,001$ coarse elements, and all together $697,570$ fine tetrahedral elements; $w=0.1$ in Eq.~\eqref{eq:target}. \emph{Infill} was conducted on two meshes respectively of $120,412$ and $286,566$ hexahedral elements. \emph{Conforming} generated a foam of $59,423$ beams (ours has $59,099$ beams).

Both ours and \emph{Conforming} generated structures of valid geometry while \emph{Infill} produced some undesired broken parts or protrusions. In addition, ours gave the best foam of the smallest compliance, demonstrating its high effectiveness. The time costs per iteration of \emph{Conforming}, \emph{Infill} and ours are respectively: 54.8s\footnote{data from~\cite{wu2019design} for a reference}, 56.9s and 396.0s. 

\textbf{Comparisons with direct FEM.}
We test the necessary of using the high-accuracy numerical coarsening during optimization by comparing performance of the resulted foams obtained via FEM on four different background meshes shown in Fig.~\ref{fig:scales}, using the femur model in Fig.~\ref{fig:teaser}. The finest is taken as reference. Each coarse element's Young's modulus is averaged from those of its interior fine elements, following a common practice. Table~\ref{tab:scales1} summarizes the results. Our method produced almost the same compliance as the reference, both of $C_{true} = 257.6$, and converged fastest in only 91 iterations. The other four mesh cases produced much worse performance and were more difficult to converge. The fact tells clearly the necessity of the CBN-based simulation. 

\begin{table}[htb]
  \centering 
  \caption{The optimization results using FEM on different meshes for the femur in Fig.~\ref{fig:teaser}: number fine elements, iterations, compliance $C$ at convergence, benchmark compliance $C_{b}$, time (in seconds) of each iteration.}
  \label{tab:scales1}
  \begin{tabular}{c|r|c|c|c|c}
  \hline
Method & \#ele\ \ \ \  & \#iter        & $C$     & $C_{b}$ & \begin{tabular}[c]{@{}c@{}}time\\ per-iter\end{tabular} \\ \hline
  FEM$_{tet}$    & 640    & -$^1$         & 173.9 & 288.2      & 9.1                                                     \\
  FEM$_{tet}$    & 5,526  & 180           & 228.9 & 265.9      & 10.3                                                    \\
  FEM$_{hex}$    & 666    & -$^1$         & 162.4 & 298.6      & 11.5                                                    \\
  FEM$_{tet}$    & 221,586 & 110          & 257.6 & 257.6  & 48.6                                                    \\
  Our    & 221,004 & 91            & 234.6 & 257.6      & 14.4                                                    \\ \hline
  \end{tabular}\\
  {\small $^1$ not converged till 200 iterations.}
\end{table}

The simulation accuracy was also evaluated by comparing simulation results on the four different meshes. Fig.~\ref{fig:simu-mesh} plots all these results. The approximation errors are $r=0.06, r=7.8e^{-3}, r=0.25, r=5.6e^{-3}$ respectively. NOTICE (JinHUANG): not consistent to the figure.
Our CBN-based simulation showed a very close approximation to the reference. It also has a much improved accuracy in comparison with other numerical homogenization approaches, as has been extensively studied in~\cite{li2022analysis}. 

\begin{figure}[!htb]
  \centering
  \includegraphics[width=0.48\textwidth]{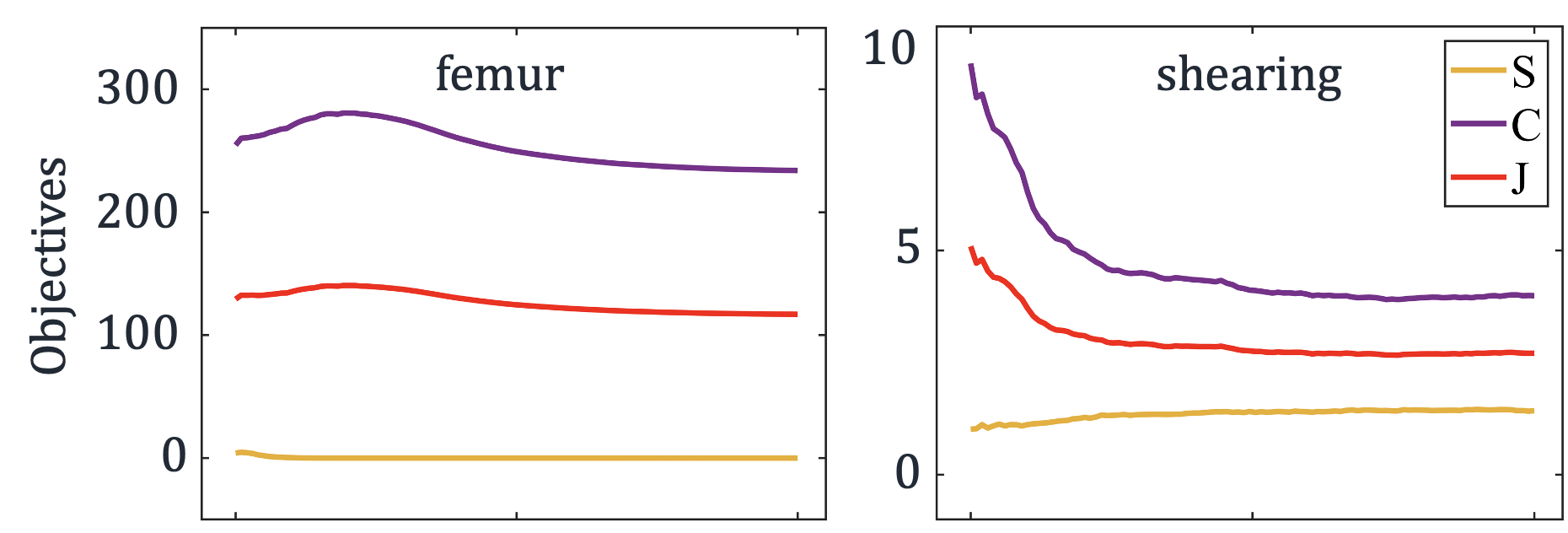}
  \caption{Variations of the compliances $C$ (purple), the shape energies $S$ (yellow), and the overall objectives $J$ (red) under two typical test cases: femur in Fig.~\ref{fig:teaser} and shearing cube in Fig.~\ref{fig:diff-v}.} 
  \label{fig:convergence}
  \centering
  \includegraphics[width=0.48\textwidth]{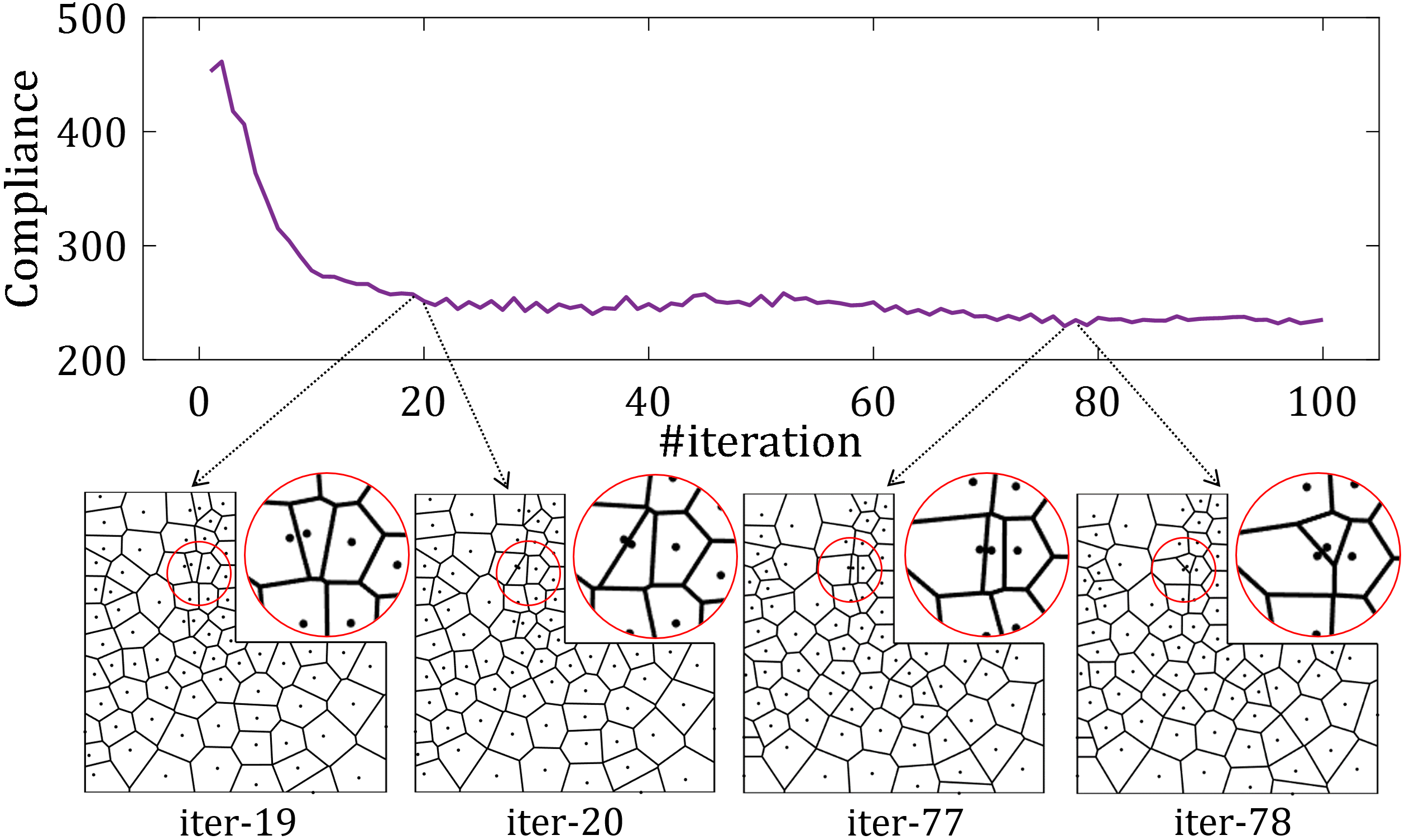}
  \caption{Convergence curve of an L-shape model and the Voronoi tessellations at some steps.}
  \label{fig:Lshape2}
\end{figure}

\subsection{Timings and convergence.}
\textbf{Timings.} The computation time is summarized in Table~\ref{tab:time}. Overall, the simulation time depends on the fine mesh resolutions, and the optimization time additionally depends on the seed number. 
Benefiting from its local approximation, the gradients were efficiently achieved. A direct Voronoi tessellations based approach would be much more expensive, for example, 6,035 seconds for the femur model in Fig.~\ref{fig:teaser}. 

\begin{table*}[t]
  \centering
  \caption{Timing (in seconds) of Voronoi tessellations (per-vor), computation of signed distance functions (per-dist), simulation (per-sim), finite difference (per-diff) in each iteration, and overall cost (per-iter) of the examples on Intel 11700 CPU. Sizes of global fine mesh, coarse mesh and the number of seeds are also summarized.}
    \begin{tabular}{c|c|r|c|r|c|c|c|c|c}
    \hline
    Model    & \begin{tabular}[c]{@{}c@{}}Fig.\\ (L/M/R)\end{tabular} & \#fine\ \ \    & \#coarse & \#seed & \begin{tabular}[c]{@{}c@{}}time\\ per-vor\end{tabular} & \begin{tabular}[c]{@{}c@{}}time\\ per-dist\end{tabular} & \begin{tabular}[c]{@{}c@{}}time\\ per-simu\end{tabular} & \begin{tabular}[c]{@{}c@{}}time\\ per-diff\end{tabular} & \begin{tabular}[c]{@{}c@{}}time\\ per-iter\end{tabular}\\ \hline
    femur   & \ref{fig:teaser}                                       & 221,004   & 640      & 500    & 2.19                                                    & 3.01                                                   & 0.71                                                    & 2.65                                                    & 8.67                                                  \\
    dome     & \ref{fig:diff-w}                                         & 206,825   & 122      & 122    & 0.33                                                    & 0.20                                                   & 8.49                                                    & 1.79                                                    & 10.99                                                  \\
    shearing & \ref{fig:diff-v} (L)                                   & 260,608   & 512      & 1,000  & 1.10                                                   & 2.17                                                   & 3.14                                                    & 3.26                                                    & 9.79                                                  \\
    insole   & \ref{fig:shoe} (L)                                     & 117,946   & 411      & 500    & 1.24                                                   & 0.92                                                   & 1.04                                                    & 1.31                                                    & 4.59                                                   \\
    \hline
    \end{tabular}
    \label{tab:time}
\end{table*}

\textbf{Convergence.}
The convergence was plotted in Fig.~\ref{fig:convergence} for two typical tests: femur in Fig.~\ref{fig:teaser} and shearing cube in Fig.~\ref{fig:diff-v}. The two cases all showed global convergences. 
To watch closely, Fig.~\ref{fig:Lshape2} plots the Voronoi variations for a concave 2D L-shape during optimization. Two pairs of exemplar cases were picked up: iter-19 to iter-20 with compliance increasing and iter-77 to iter-78 with compliance decreasing. Drastic cell topology variations were observed for both cases, which may cause inaccurate gradient computations (see also Sec.~\ref{sec-diff-num}) and consequently the un-smooth convergence. Still, an optimized Voronoi foam was robustly obtained.

\subsection{Influence of parameter selections}
 \begin{figure}[!htb]
   \centering
    \includegraphics[width=0.45\textwidth]{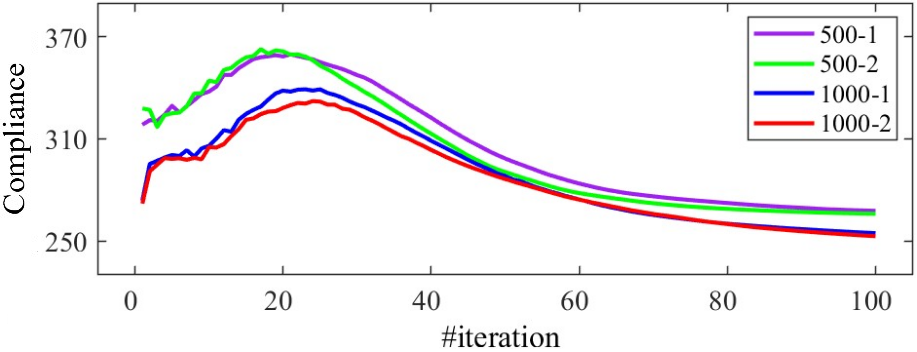}
   \caption{Convergence curves under four different initial seeds, for the femur model in Fig.~\ref{fig:teaser}. Two (500-1, 500-2) contain 500 seeds with different positions, and another two (1000-1, 1000-2) contain 1000 seeds with different positions.}
  \label{fig:seeds_curve}
  \includegraphics[width=0.48\textwidth]{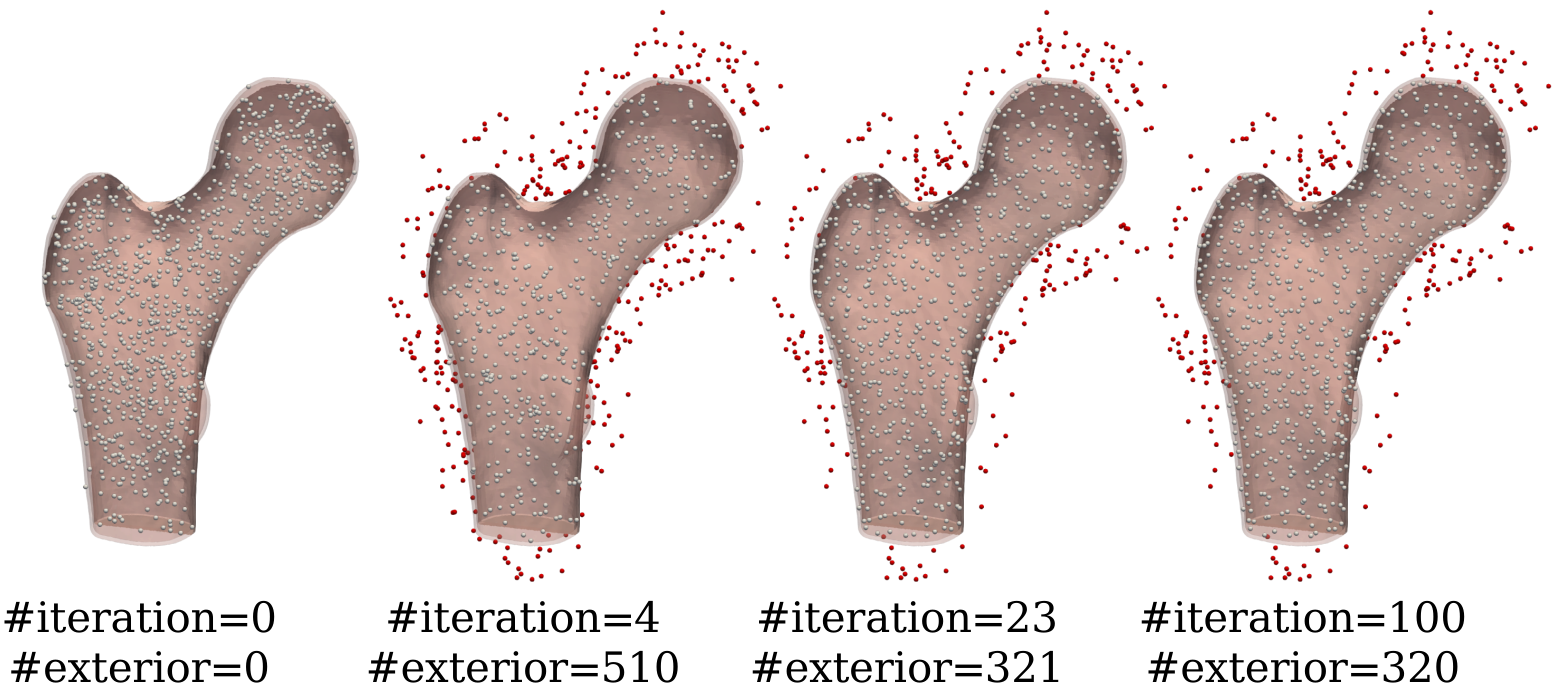}
  \caption{Seed positions in four iterations for a case of 1000 seeds, where the white seeds are interior and the red are exterior. The exterior seeds did not contribute to the optimization model. 
  }
  \label{fig:seeds_four_iterations}
 \end{figure}

\textbf{Different initial seeds.}
We tested the method's adaptivity to the numbers and positions of initial seeds using the femur model in Fig.~\ref{fig:teaser} for four different seed sets: of random 500 seeds (500-1, 500-2) and of 1000 seeds (1000-1, 1000-2). As shown in Fig.~\ref{fig:seeds_curve}, very close compliance was observed for cases of the same amount of seeds, illustrating the independence of initial seed positions. Slightly stiffer foams (with smaller compliance) were generated with 1000 seeds. 


The approach also demonstrated its capability in adjusting the seed number by automatically moving unnecessary seeds outside of the outer shape. The seed movements were plotted in Fig.~\ref{fig:seeds_four_iterations}. 
Consider the 1000-1 case. $320$ seeds out of 1000 contributed to the final Voronoi foam, with a volume fraction from $0.31$ to $0.25$.  Benefiting from this, a rough estimate of the amount of seeds is sufficient for the users, avoiding multiple tedious attempts. 

\begin{figure}[!htb]
  \centering
  \includegraphics[width=0.48\textwidth]{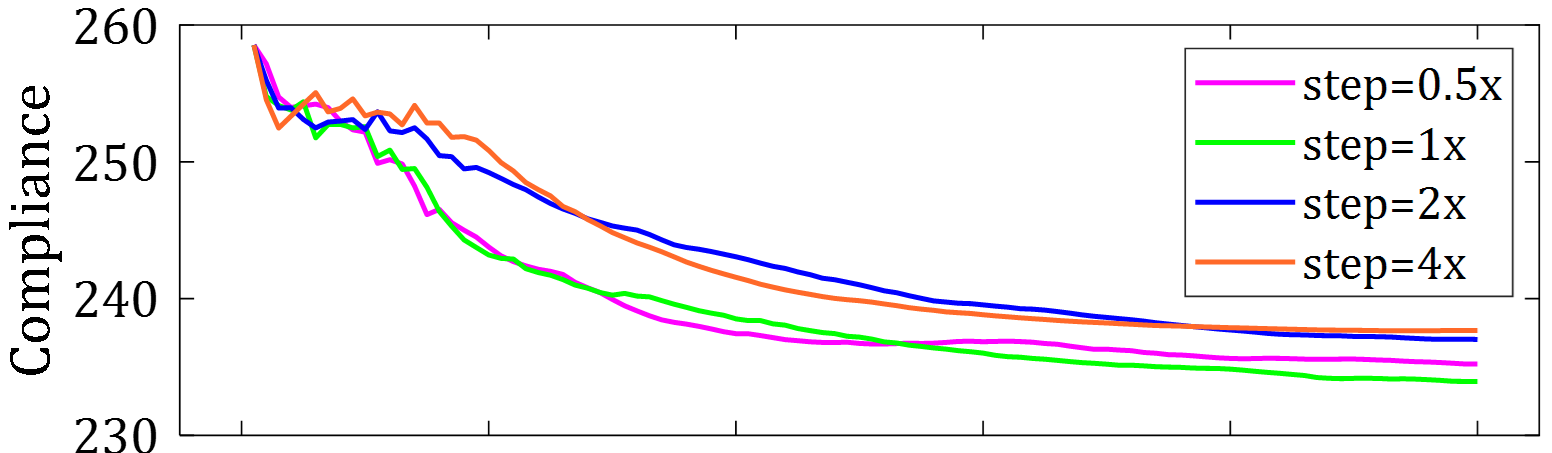}
  \caption{Convergence curves of finite differences under four different steps, of 0.5, 1, 2, 4 times the average edge length of fine elements, for the femur model in Fig.~\ref{fig:teaser}.}
  \label{fig:diff2}
\end{figure}

\textbf{Different finite difference steps.} We tested the method's convergence and stability under different finite-difference steps for the femur model at different steps: 0.5x, 1x, 2x, 4x the average side length of fine elements. The convergence curves of the compliance $C$ and the shape energy $S$ in Fig.~\ref{fig:diff2} showed an overall convergence. The case of step=1x gave the stiffest foam and was set as default. The shape energy showed a more stable and smooth convergence than the compliance as the former was computed analytically. 

%


\subsection{Practical applications in different cases}



\begin{figure*}[!htb]
  \centering
  \includegraphics[width=0.8\textwidth]{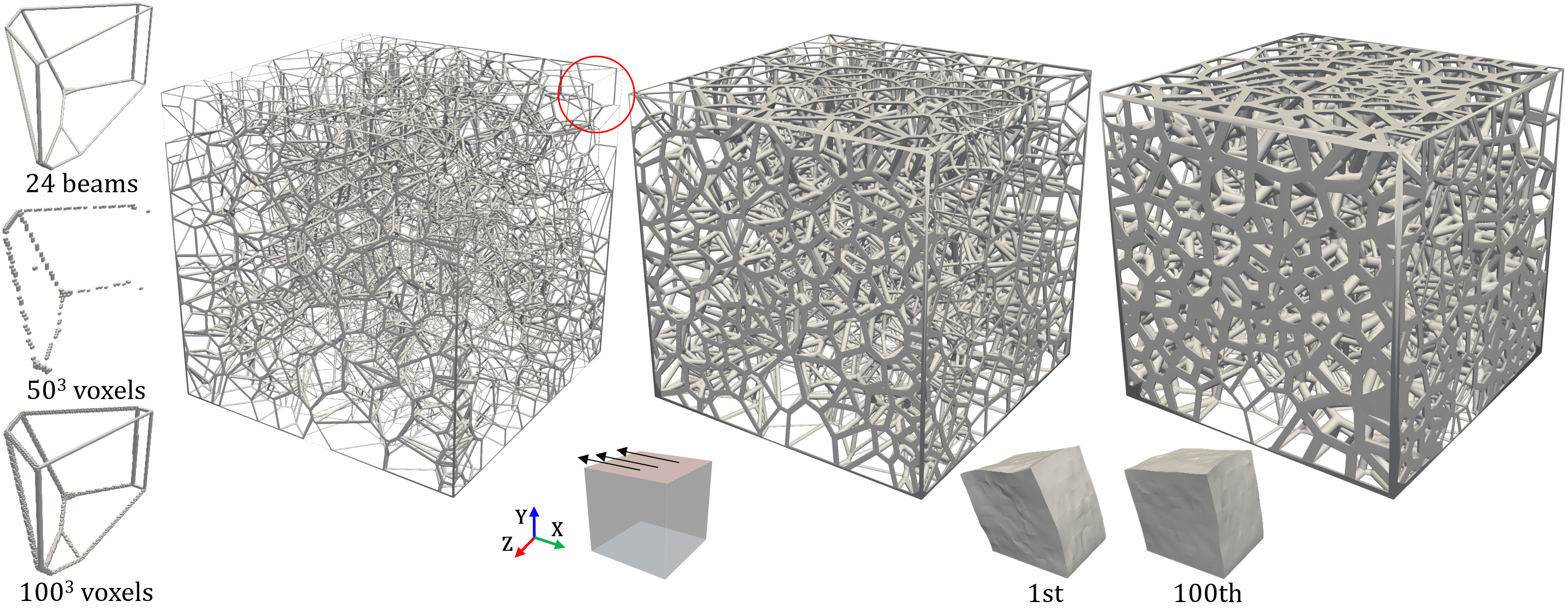}
  \caption{Different volume fractions of extremely small values $v = 0.02, 0.05, 0.1$. Under the shearing boundary conditions, the deformations of the foams with $v = 0.02$ at 1st iteration and 100th iteration are plotted at the bottom, respectively of compliances $7.6$ and $4.4$. The left plots the voxel-based representations for a typical Voronoi cell at different resolutions.}
  \label{fig:diff-v}
\end{figure*}

\textbf{Extreme low volume fractions.} 
We tested the approach's capacity under extremely low volume fractions $v= 0.02,0.05,0.1$ using the shearing cube in Fig.~\ref{fig:diff-v}. The case is very challenging for voxel-based topology optimization~\cite{andreassen2011efficient}. 
%
If being represented as voxels as in voxel-based topology optimization, each foam cell needs approximately $100^3$ voxels to capture the details and the overall foam needs around $1$-billion voxels. It would be too computationally expensive, not to mention its difficulty in valid geometry control. Our approach only needs $4,000$ design variables for the optimization. 

\begin{figure}[!htb]
  \centering
  \includegraphics[width=0.48\textwidth]{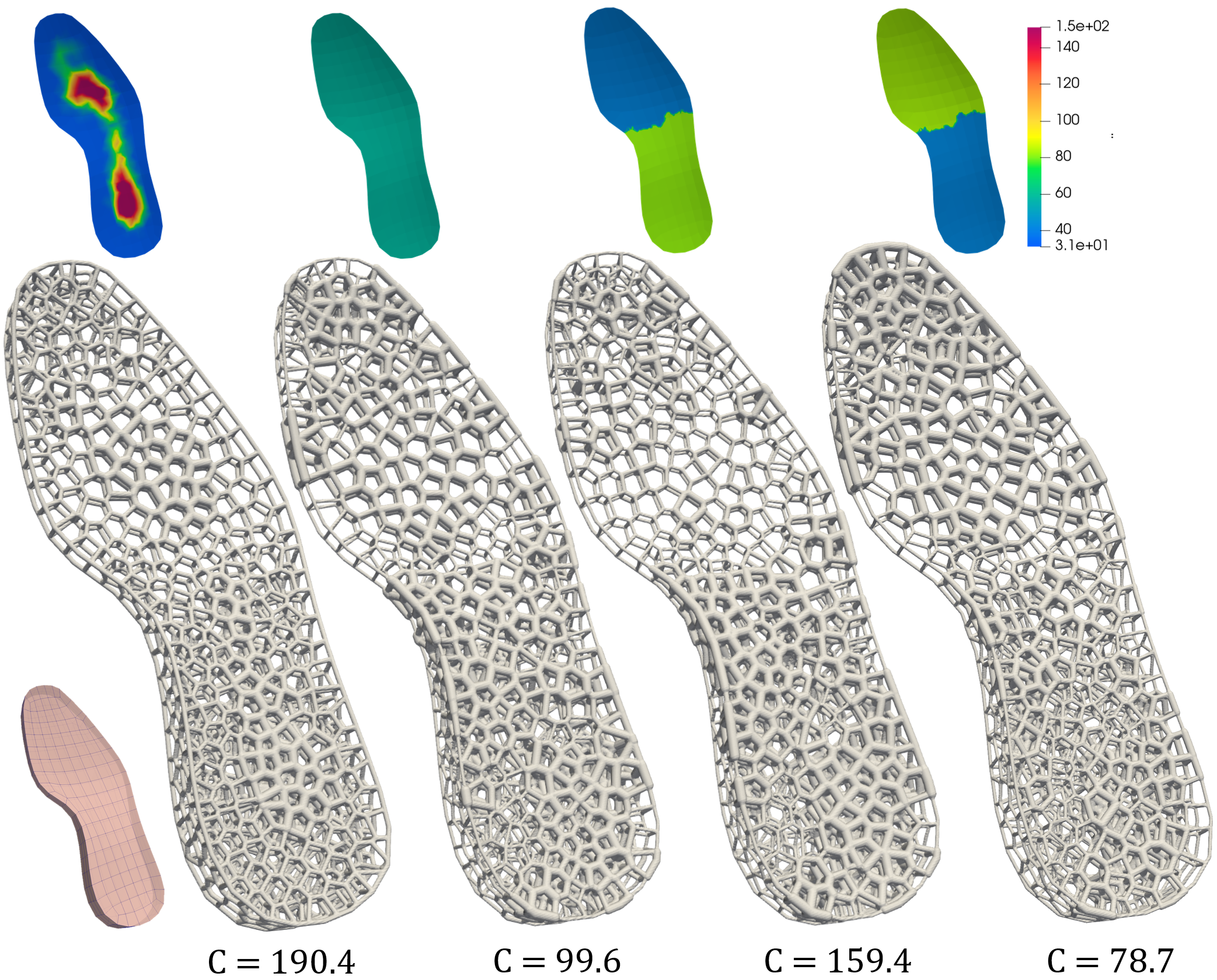}
  \caption{Different loading forces, including mimicking real foot pressure, constant pressure, constant pressure of 3x higher distribution in the heel or the front. The compliances have been scaled $1e^{-6}$ times.}
  \label{fig:shoe}
\end{figure}

\textbf{Different loading forces.} Results were shown in Fig.~\ref{fig:shoe} on an insole model at three different loadings: real foot pressure, constant pressure, constant pressure of 3x higher distribution in the heel or the front~\cite{xu2015interactive}. 
Denser seed distributions and larger radii appeared in the larger pressure area to maintain stronger stiffness.

\begin{figure}[!htb]
  \centering
  \includegraphics[width=0.42\textwidth]{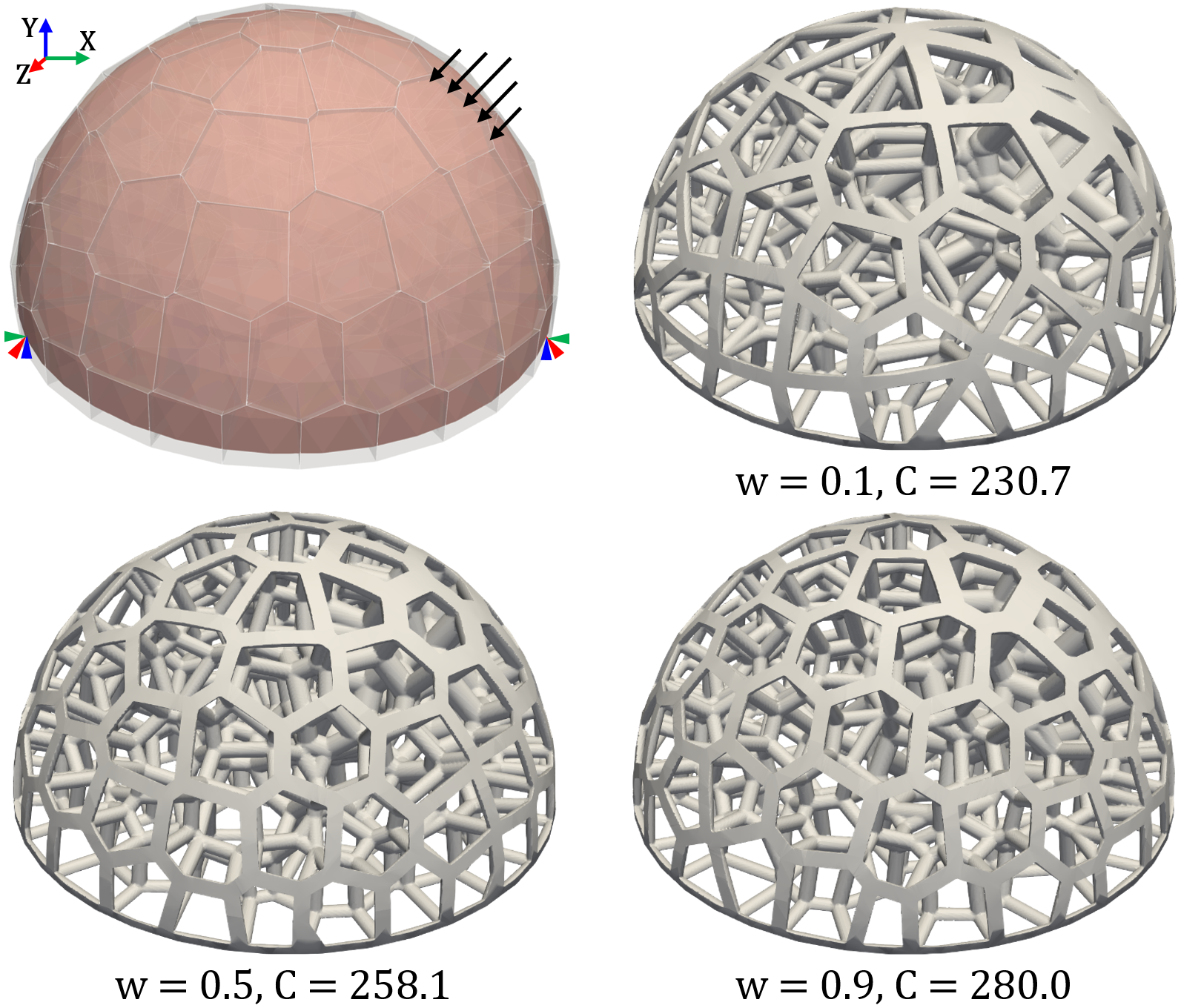}
  \caption{Different shape regularization weights $w=0.1, 0.5, 0.9$. The main bearing beams with smaller $w$ tend to be thicker than those with bigger $w$.}
  \label{fig:diff-w}
\end{figure}

\textbf{Shape regularization weights.} 
The shape regularization weight $w$ in Eq.~\eqref{eq:prob} was set of three different values $w = 0.1,0.5,0.9$, on a dome model of coarse polyhedral mesh in Fig.~\ref{fig:diff-w}. The smaller $w$, the stiffer model with smaller compliance and worse shape regularization were generated. This is consist with our heuristics. 
The dome model took more computation time, perhaps because its denser stiffness matrix from its polyhedral background mesh. 

\begin{figure}[!htb]
  \centering
  \includegraphics[width=0.25\textwidth]{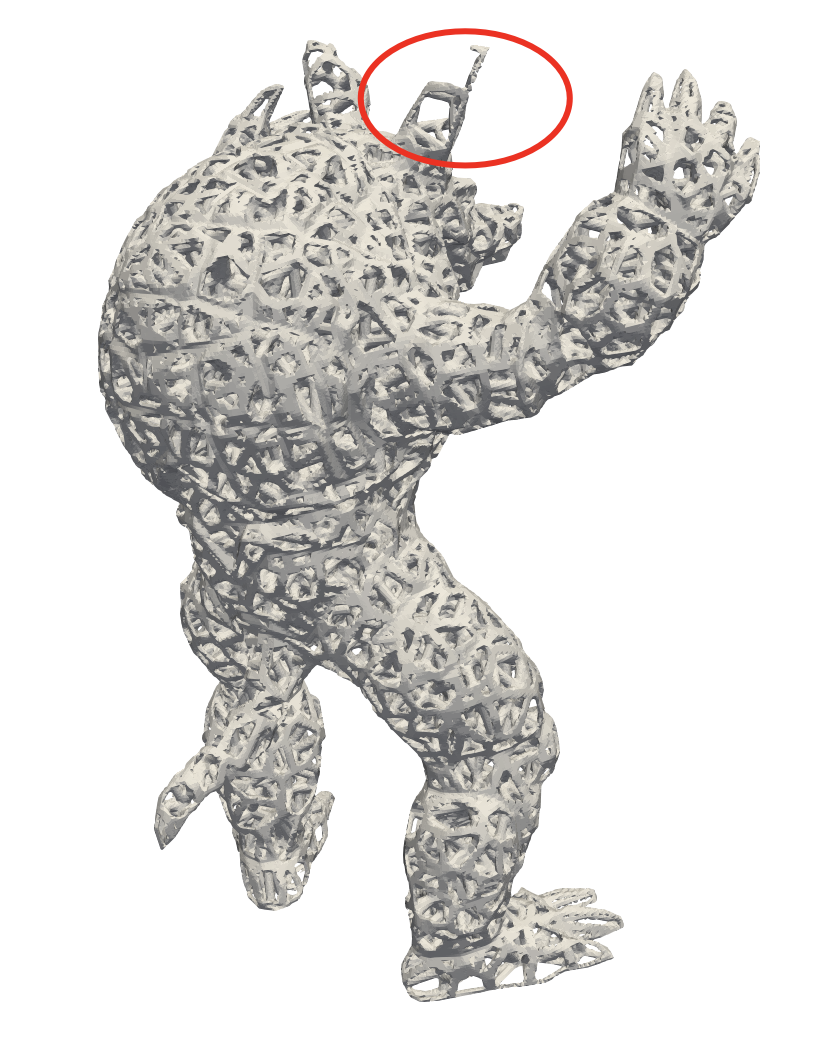}
  \caption{Our approach may produce a geometrically invalid foam around extremely cusp corners.}
  \label{fig:fail}
\end{figure}

\textbf{Failure case}. Our approach is generally able to produce geometrically valid Voronoi foams for free-form shapes. It may fail for shapes of extremely cusp corners because of the boundary clipping. See for example the right ear region in the Armadillo in Fig.~\ref{fig:fail}. The issue is to be further explored in our future work. 

\section{Conclusion}\label{sec-conclusion}
We propose an explicit topology optimization method for conforming
open-cell foam design using Voronoi tessellation. Its usage of synchronized explicit and implicit representation in modeling, simulation and optimization
offers unique advantages on reliable and efficient Voronoi foam
optimization. It also answers two general critical technical questions
in implementing the goal on efficient gradient computation and reliable 
property simulation. 
The approach is always being able to produce a geometrically valid foam structure, even at volume fraction as low as $2\%$, which is never observed in conventional voxel-based topology optimization. 

The approach opens a new avenue for reliable topology optimization of porous foams by always maintaining its geometric validity, resolving the open question in topology optimization~\cite{Wu2021Review}. Noticing that any 2D triangulation can be represented through a perturbation of a weighted Delaunay triangulation, a dual form of Voronoi tesselation ~\cite{Memari2011Para}. The approach may thus be of great generality in producing general open-cell foams. The topic is to be explored in our future work. At present, it at least can be extended as follows. First, we are to devise fully analytical derivatives for a more stable and efficient of Voronoi foam optimization. Second, the open-cell foam has distinguishing property of impact absorption than a close-cell one. Extending the approach for the associated  topology optimization is to be studied, which must account for the nonlinear large deformations. Third, we will explore approaches in introducing anisotropy~\cite{du1999centroidal, Lvy2010LpCV, Budninskiy2016OptimalVT, Fang2020ReinforcedF} into the Voronoi foam to improve its performance.  

\section*{Acknowledgments}
We would like to thank all the anonymous reviewers for their valuable comments and suggestions. The work described in this paper is partially supported by the National Key Research and Development Program of China (No. 2020YFC2201303), the NSF of China (No. 61872320), the Key R\&D Program of Zhejiang Province (No. 2022C01025), and the Zhejiang Provincial Science and Technology Program in China (No. 2021C01108).

\bibliographystyle{IEEEtran}
\bibliography{AllBibTex-v17}

\begin{IEEEbiography}[{\includegraphics[width=1in,height=1.25in,clip,keepaspectratio]{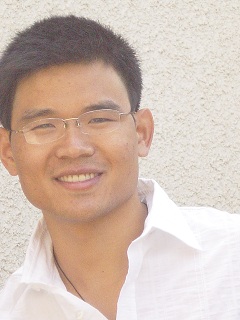}}]{Ming Li}
is an associate professor at the State Key Laboratory of CAD\&CG, Zhejiang University, China. He received his PhD degree in Applied Mathematics from the Chinese Academy of Sciences in 2004. From that on, he worked in Cardiff University, UK, and Drexel University, USA for four years. His research interest includes CAD/CAE integration, porous foams and design optimization. Dr. Li has published more than 50 peer-reviewed international papers on the topics.
\end{IEEEbiography}

\begin{IEEEbiography}[{\includegraphics[width=1in,height=1.25in,clip,keepaspectratio]{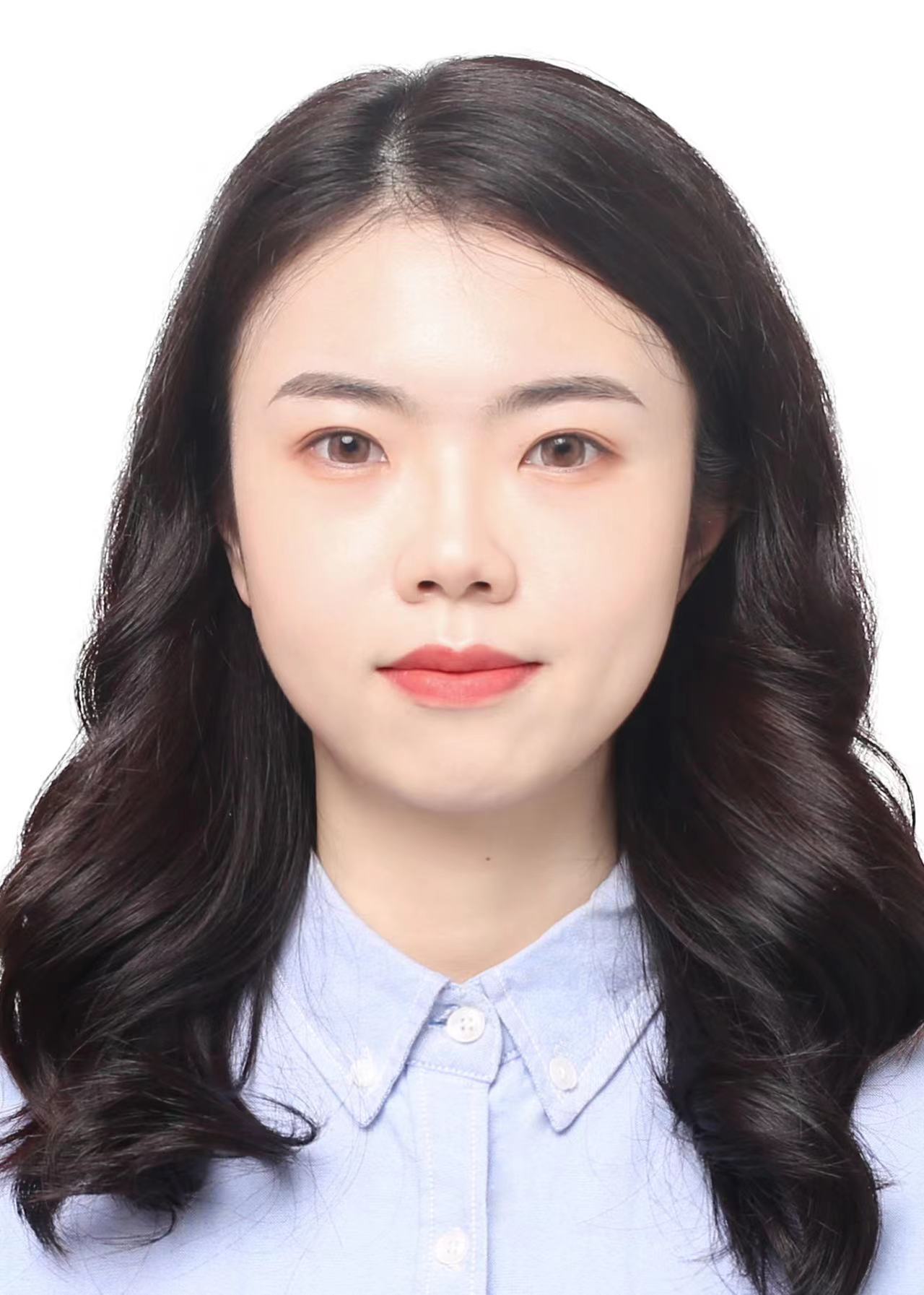}}]{Jingqiao Hu}
received the PhD degree in Computer Science from Zhejiang University, China, in 2022. Her research interests include numerical coarsening and topology optimization. 
\end{IEEEbiography}

\begin{IEEEbiography}[{\includegraphics[width=1in,height=1.25in,clip,keepaspectratio]{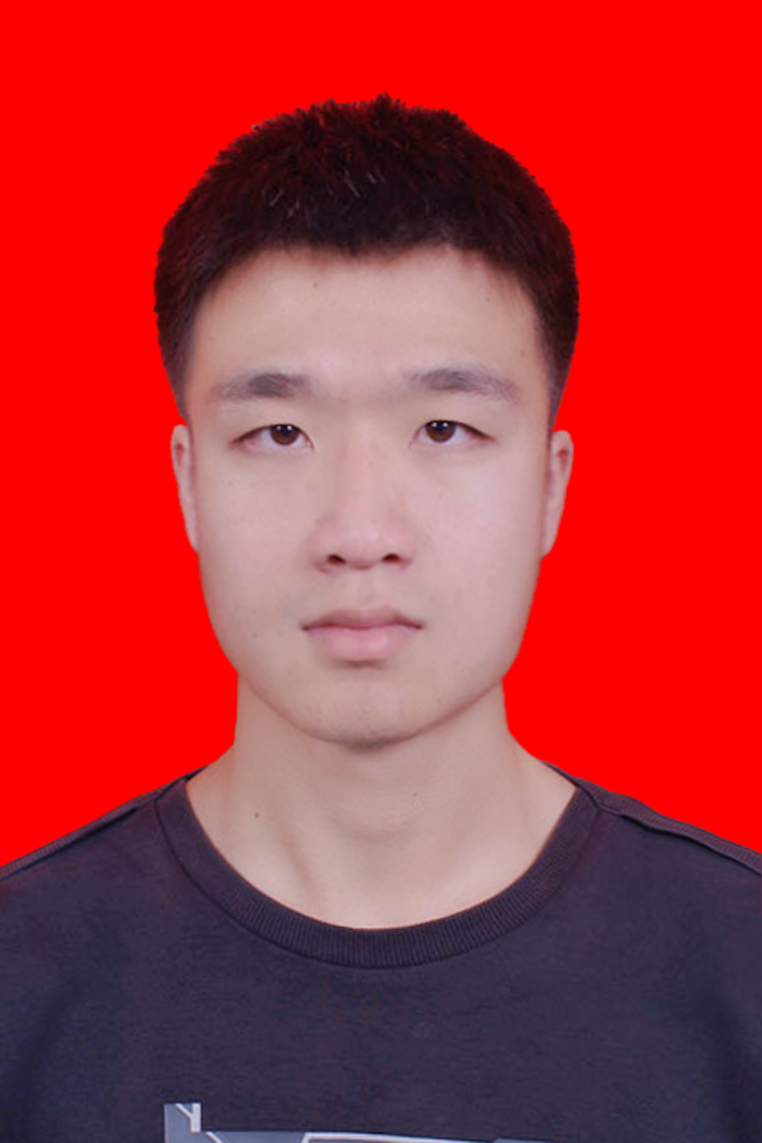}}]{Wei Chen}
received the bachelor's degree in Computer Science from Lanzhou University, China, in 2021. He is now a PhD student in the State Key Laboratory of CAD\&CG, Zhejiang University. His research interests include physical simulation and computer-aided design.
\end{IEEEbiography}

\begin{IEEEbiography}[{\includegraphics[width=1in,height=1.25in,clip,keepaspectratio]{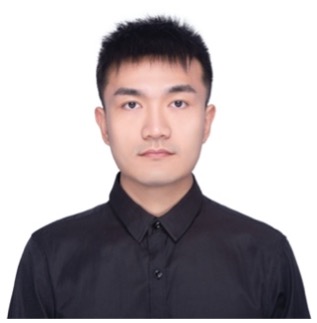}}]{Weipeng Kong}
received the Master's degree in Computer Science from Zhejiang University, China, in 2023. His research interests include computer graphics and architectures of CAD system.
\end{IEEEbiography}

\begin{IEEEbiography}[{\includegraphics[width=1in,height=1.25in,clip,keepaspectratio]{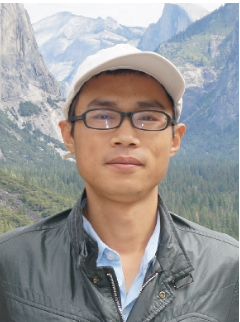}}]{Jin Huang}
is a professor in the State Key Lab of CAD\&CG of Zhejiang University, China. He received his PhD. degree in Computer Science Department from Zhejiang University in 2007 with Excellent Doctoral Dissertation Award of China Computer Federation. His research interests include geometry processing and physically-based simulation. He has served as reviewer for ACM SIGGRAPH, EuroGraphics, Pacific Graphics, TVCG etc.

\end{IEEEbiography}

\end{document}